\documentclass[%
 reprint,
nofootinbib,
 amsmath,amssymb,
 aps
]{revtex4-1}
\pdfoutput=1
\usepackage{graphicx}% Include figure files
\usepackage[utf8]{inputenc}
\usepackage{dcolumn}% Align table columns on decimal point
\usepackage{bm}% bold math
\usepackage{verbatim}
\usepackage{color,ulem}

\usepackage[utf8]{inputenc}

\def\be{\begin{equation}}
\def\ee{\end{equation}}
\def\bea{\begin{eqnarray}}
\def\eea{\end{eqnarray}}

\begin{document}

\preprint{}[IFT-UAM/CSIC-19-116]

\title{\huge 
Scale Invariant Solids
}% Force line breaks with \\
%\thanks{A footnote to the article title}%

\author{Matteo Baggioli}%
 \email{matteo.baggioli@uam.es}
 \thanks{\url{https://members.ift.uam-csic.es/matteo.baggioli/} }
\affiliation{Instituto de Fisica Teorica UAM/CSIC,
c/ Nicolas Cabrera 13-15, Cantoblanco, 28049 Madrid, Spain
}%

 \author{Víctor Cáncer Castillo}%
 \email{vcancer@ifae.es}
 \author{Oriol Pujol{\`a}s}%
 \email{pujolas@ifae.es}
\affiliation{Institut de F\'isica d'Altes Energies (IFAE), The Barcelona Institute of Science and Technology (BIST)\\
Campus UAB, 08193 Bellaterra, Barcelona.
}%

\begin{abstract}

Scale invariance (SI) can in principle be realized in the elastic response of solid materials. 
There are two basic options: that SI is a manifest symmetry or that it is spontaneously broken.
The manifest case corresponds physically to the existence of a non-trivial infrared fixed point with phonons among its degrees of freedom.
We use simple bottom-up AdS/CFT constructions to model this case. We characterize the types of possible elastic response
and discuss how the sound speeds can be realistic, that is, sufficiently small compared to the speed of light. 
We also study the spontaneously broken case using Effective Field Theory (EFT) methods. 
We present a new one-parameter family of nontrivial EFTs that includes the previously known `conformal solid' as a particular case 
as well as others which display small sound speeds.
We also point out that an emergent Lorentz invariance at low energies could affect by order-one factors the relation between sound speeds and elastic moduli.

\end{abstract}

\pacs{Valid PACS appear here}
\maketitle

\section{Introduction}

The mechanical response of matter under small applied stresses is a well-known subject \cite{landau7,Lubensky}. At sufficiently low energies, it can be described in a continuum limit by the so-called elasticity theory. 
Just like in hydrodynamics, the main assumption is that the displacements in the solid are described by an {\rm effective} set of fields $\phi^i(t,x^j)$ that represent the deformations of the material from its equilibrium position at each point. The effective Lagrangian for $\phi^i(t,x^j)$ is then automatically fixed by symmetries. It was shown in \cite{Leutwyler:1996er} (see also \cite{Dubovsky:2005xd}) that the form of the nonlinearities in the effective Lagrangian is greatly constrained by the fact that the phonon field $\phi^i(t,x^j)$ can be viewed as the Goldstone boson arising from the spontaneous breaking of translation invariance. More recently, it has been shown how to derive the effective Lagrangian applying the Coset construction to the spontaneous breaking of Poincar{\'e} symmetry  \cite{Nicolis:2013lma,Nicolis:2015sra}. 

These developments taught us how to promote elasticity theory into a fully nonlinear Effective Field Theory (EFT). 
We shall refer to this EFT simply  as {\it Solid EFT} and give more details on it below.
As it happens with other known  EFTs, one expects that this provides for an efficient way to re-sum 
certain low-energy observables that are difficult to compute directly from the microscopic theory. 
It is natural, then, to ask what are the phenomenological consequences that can be extracted and how the procedure works. Given that the EFT methods mainly provide nontrivial information concerning the nonlinear part of the theory, one expects that the Solid EFT provides interesting constraints/information about the phonon interactions (e.g., phonon $2\to2$ scattering), but more generally also regarding the nonlinear elastic response. Ref.~\cite{Alberte:2018doe}, initiated a study in this direction, showing that nontrivial relations among several nonlinear observables can indeed be identified.  This motivates us to continue the analysis to more sophisticated cases.

The purpose of this work is to focus on the special case where the solid exhibits {\it scale invariance} (SI), in addition to  the broken symmetries of a regular solid. Aside from being interesting {\it per se}, this case seems to be quite close to real world of materials that exhibit criticality in the form of a quantum critical point. 
In order to possibly make contact with these especially interesting  materials, it is desirable to understand well how SI is compatible with solid EFT or similar techniques.\footnote{In this work, SI is meant to be realized in the mechanical sector -- by the phonons. 
It isn't our goal to identify what kind of physical system accomplishes this, but the idea is very well posed so we just take it as an assumption.}

\begin{figure}
\centering
\includegraphics[width=\linewidth]{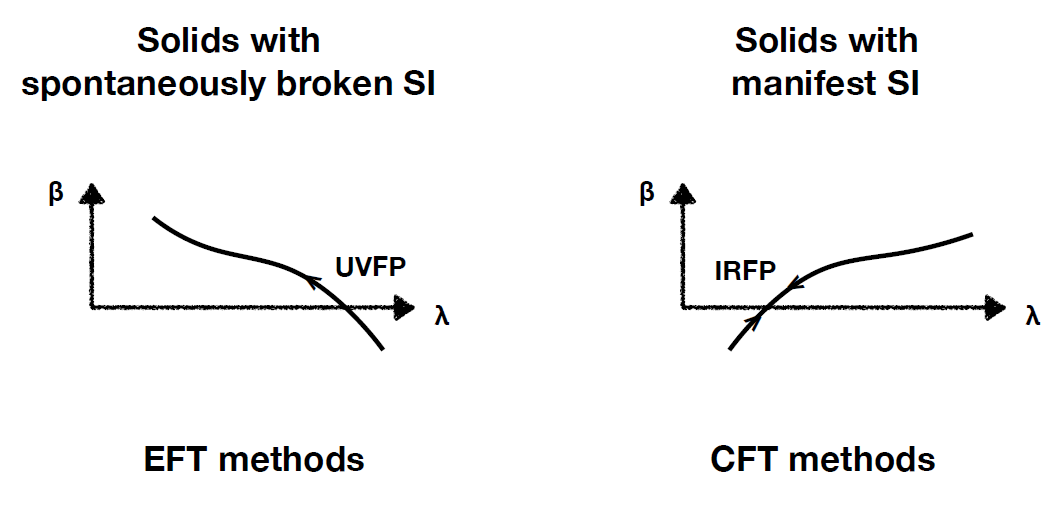}
\caption{Sketch of the two basic options on how scale invariance (SI) can be realized, in terms of the beta function $\beta=\mu \frac{d\lambda}{d\mu}$ of a certain coupling $\lambda$. The arrows indicate the flow towards low energies. The left plot represents the spontaneous breaking case, which we discuss in Section 3. The right cartoon represents the case with manifest SI, which we discuss in Section 4.
}
\label{options}
\end{figure}

It is worth spending some words on what are the possible ways how SI can be realized generically. 
Conceptually, the main division arises from whether or not the low energy dynamics is governed by a nontrivial infrared fixed point (IRFP). 
To some extent, in the presence of an IRFP one can say that SI is an unbroken symmetry. By the same logic, in the absence of an IRFP, then, SI can only be spontaneously broken -- a nonlinearly realized symmetry. This criterion allows to separate possible realizations of SI in solids in two basic groups:
\begin{itemize}
\item  {\bf solids with spontaneously broken SI}. In this case one expects a gapped spectrum and that the phonons are isolated degrees of freedom at low energies. In this case EFT methods are applicable in order to describe the lightest excitations (the phonons) in the mechanical sector as the Goldstone bosons of the spontaneously broken spacetime symmetries.  We discuss this case in Section 3.
%We shall refer to the resulting EFTs as {\bf SIS-EFTs}
\item {\bf  solids with manifest SI}, where by assumption there is a dynamical IRFP. %in addition to the SB of SI. 
In this case the phonons are not isolated degrees of freedom and one expects that the dispersion relation develops an imaginary part. 
Will use bottom-up AdS/CFT methods, which are well suited to construct simple models with these properties, in Section 4.  
\end{itemize}

These  options can be better visualized using the renormalization group (RG) language, that is, in terms of
 the beta function $\beta=\mu \frac{d\lambda}{d\mu}$ of a certain coupling $\lambda$. 
The main two options are depicted  in Fig.~\ref{options}. 
The spontaneously broken SI case can be viewed as a departure (an RG-flow) from a UVFP  induced by the vacuum expectation value of some operator. 
The manifest SI case corresponds to the presence of an infrared or {\it emergent} fixed point.

The pictures in Fig.~1  also immediately suggest that one can construct more options by `combining' the two possibilities, that is by having both an IR and a UV fixed points.  For instance, one can break spontaneously the UV SI but then `land' on an IRFP which realizes an {\it emergent} SI. This case would combine both spontaneously broken and manifest realizations.

These possibilities seem to apply both to Lorentz invariant and non-invarant situations, and one can easily construct examples in bottom-up holographic models. For instance, Lorentz-invariant examples of the SB case can be found in  \cite{Bianchi:2001de} (see also \cite{Megias:2014iwa}) and of the {\it emergent + SB} case in \cite{Hoyos:2013gma,Bajc:2013wha}.

An important qualitative distinction in the Lorentz invariant case is that SB of scale (and conformal) invariance is accompanied with the appearance of a massless dilaton. It is well known that this requires fine-tuning of the theory, however assuming the tuning, the dilaton pole must appear and has indeed been found in both SB \cite{Bianchi:2001de,Megias:2014iwa} and {\it emergent+SB} cases \cite{Hoyos:2013gma,Bajc:2013wha}. 
In condensed matter setups, however, Lorentz boosts are broken and the dilaton does not appear even if SI is broken spontaneously  \cite{Low:2001bw,Esposito:2017qpj} -- in a sense it is replaced by other Goldstone bosons, the phonons. 
This motivates a deeper study of the possible realizations of SI in solid materials from the low energies effective point of view, with the main focus in whether SI is a spontaneously broken or a manifest symmetry.

The second main motivation for our work is, perhaps, more down-to-earth: a {\it sine qua non} condition for the theories that aim to describe realistic solids (SI or not) is that the sound speeds (both transverse and longitudinal) are tiny compared with the speed of light $c$ -- indeed, the fastest sound speeds in known materials are around  $10^{-4}$ in units of $c$. 
To the best of our knowledge, the only know previous example of a SI solid effective theory is the so-called conformal solid \cite{Esposito:2017qpj}, and it displays inevitably relativistic longitudinal sound waves. Any effective description of realistic  SI solids must  overcome this difficulty. 
Below, we will show how SI is compatible with `slow sound'  both for spontaneously broken and manifest SI.

The mechanical response in critical materials is also the subject of recent research from a more condensed matter perspective \cite{ishii2019glass,setty2019glass,Sahu_2019}. They are also of interest since it is in this class of material that deviations from the KSS viscosity bound \cite{Kovtun:2004de} have been identified \cite{Hartnoll:2016tri,Alberte:2016xja,Burikham:2016roo,2018arXiv180108627G}.

The rest of this work is organized as follows. In Section 2.1 we review the text-book elasticity theory, and in Section 2.2 we review how it is reformulated in the EFT of solids. In Section 3 we discuss the special case of solid EFTs that incorporate SI (which corresponds to the spontaneous breaking case). In Section 4 we discuss case with manifest SI, and we conclude in Section 5.

\section{Linear elasticity}\label{sec:el}

In this Section we review a few basic notions on the elasticity theory for general (isotropic and homogeneous) solids. The main focus will be to introduce the shear and bulk (or compressibility) moduli, and their relation with the sound/phonon properties.

\subsection{Elastic response}\label{sec:linel}

\begin{figure}
\centering
\includegraphics[width=0.7\linewidth]{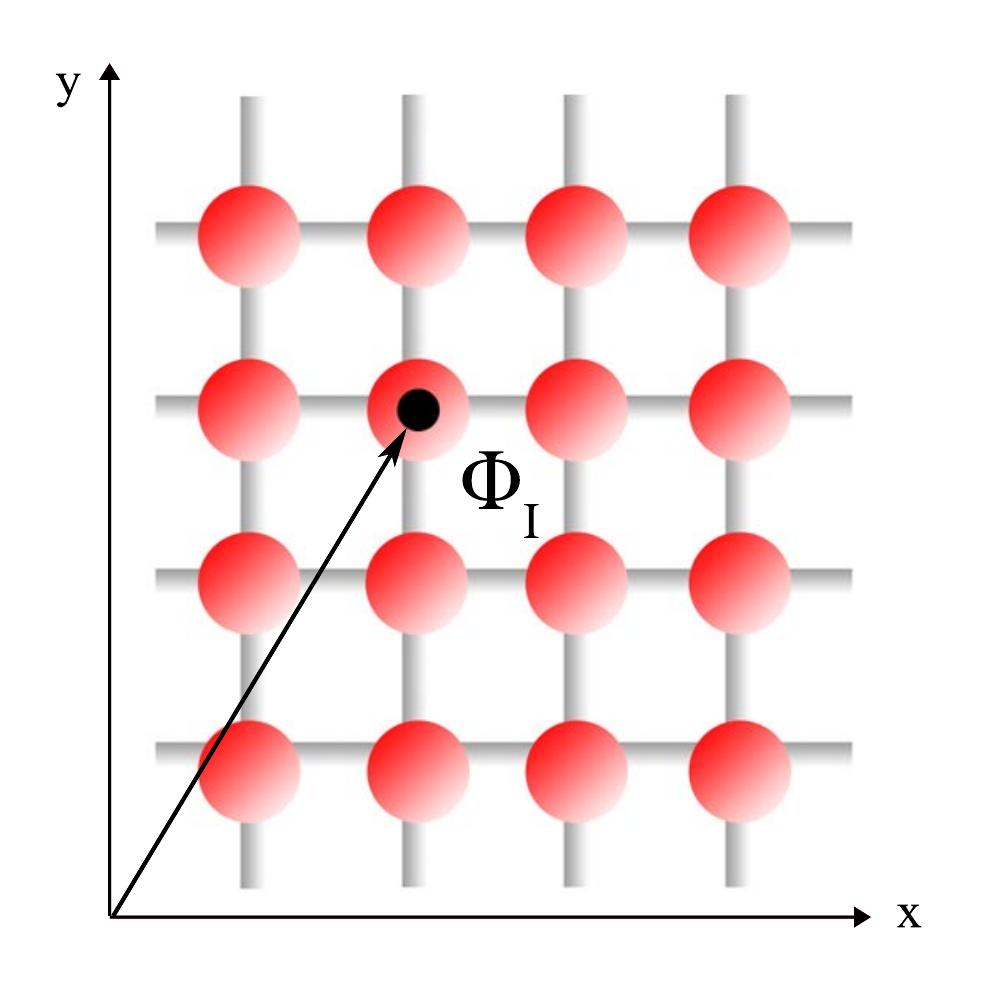}
\quad
\includegraphics[width=0.7\linewidth]{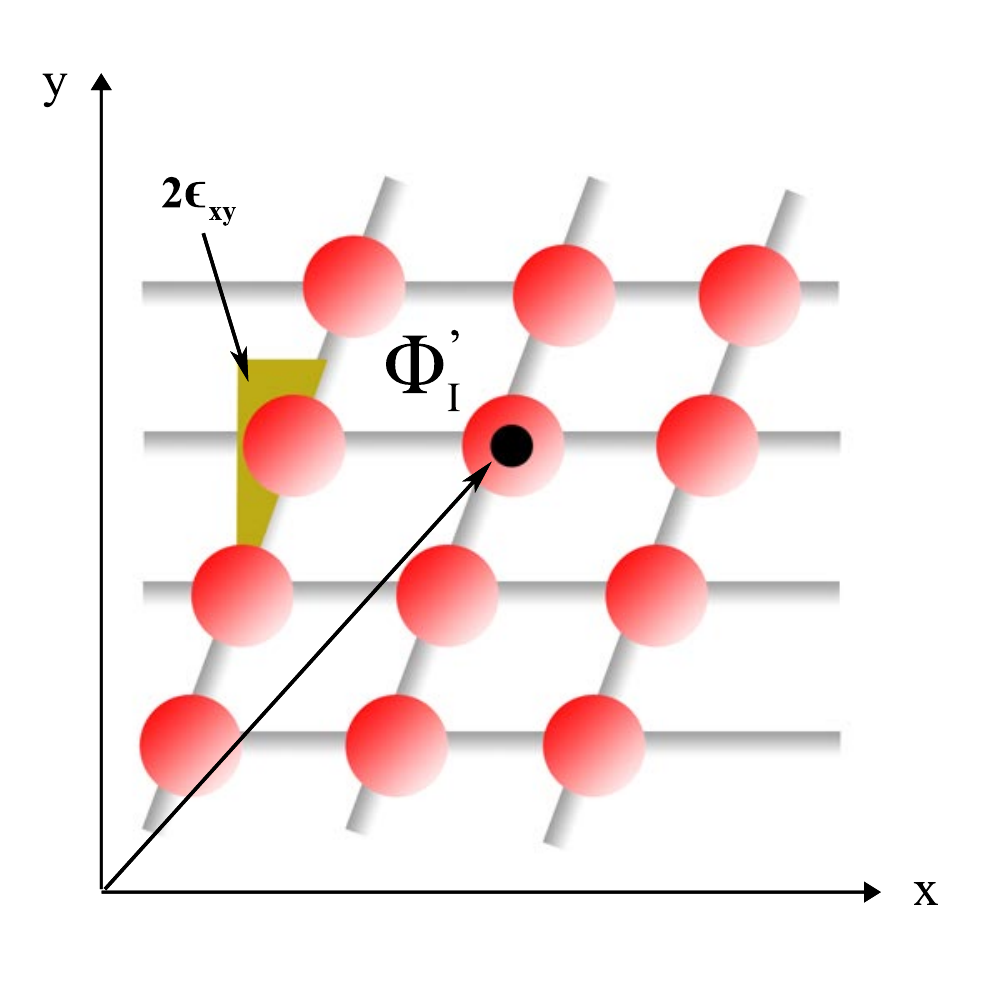}
\caption{The description of the mechanical deformations in terms of the displacement vector $\Phi^i=x^i+\phi^i$. \textbf{Top: }The equilibrium configuration is simply $\Phi^I_{eq}=x^I$. \textbf{Bottom: }An example of a shear deformation and the geometrical interpretation of the strain tensor $\varepsilon_{ij}$. The configuration changes from $\Phi^i_{eq}$ to $\Phi^i$ such that $\Phi^i-\Phi^i_{eq}= (\varepsilon_{xx}\,dx\,+\,\varepsilon_{xy}\,dy,\varepsilon_{yy}\,dy\,+\,\varepsilon_{xy}\,dx)$.}
\label{figsketch}
\end{figure}

The elastic response describes the produced \textit{stress} in a material with respect to an external mechanical deformation, \textit{i.e.} the \textit{strain} \cite{Lubensky}.
The state of mechanical deformation of the solid can be described by the mapping
\be\label{phiphi}
\Phi^i = x^i + \phi^i(t,x)
\ee
which gives the position of every given solid element. The deviations from equilibrium are directly encoded in  $\phi^i$, which act in all respects like a set of dynamical scalar fields. Their (small) wave excitations are the phonons. And the time-constant configurations of the form $\phi_i \propto x^i$ encode small shear or bulk strain deformations. 
The useful way to parametrize them is the tensorial object known as \textit{strain tensor}:
\begin{equation}
\varepsilon_{ij}\,=\,\frac{1}{2}\left(\partial_i \phi_j\,+\,\partial_j \phi_i\right)
\end{equation}
where $\phi_i\equiv r_i'-r_i$ is the displacement vector, \textit{i.e.} the deformation from equilibrium (see Fig.~\ref{figsketch}). 
The \textit{bulk strain} is defined as the trace of the strain tensor,
\begin{equation}
\varepsilon_{ii}\,=\,\vec{\partial} \cdot \vec{\phi}
\end{equation}
and it can be either positive or negative. It physically corresponds to an external compression/traction on the system which changes the volume of the sample. The \textit{shear strain} $\varepsilon$ is on the contrary the traceless part, which can be reduced to the off-diagonal component of the strain tensor,
\begin{equation}
\varepsilon\,\equiv\,2\,\varepsilon_{xy}
\end{equation}
and it encodes the angular deflection of a point from its original position (see Fig.~\ref{figsketch}). Obviously a generic mechanical deformation contains a superposition of a bulk and shear strain.

For homogeneous and isotropic solids, the response to small external strain is described at linear level by two elastic moduli. In the linear response approximation, \textit{i.e.} for small deformations, 
the deviation of the stress tensor from equilibrium
$$\sigma_{ij} \equiv T_{ij} - p\; \delta_{ij}$$
for an isotropic solid (see textbooks \cite{Lubensky,landau7}), in $d$ space dimensions, takes the simple form:
\begin{equation}
\sigma_{ij}\,=\,\mathcal{K}\,\delta_{ij}\,\varepsilon_{kk}\,+\,2\,\mathcal{G}\,\left(\varepsilon_{ij}-\,\frac{1}{d-1}\,\delta_{ij}\,\varepsilon_{kk}\right)
\label{Tlin}
\end{equation}
where  $\mathcal{G},\mathcal{K}$ are referred to as the linear shear and bulk elastic moduli respectively.

The shear elastic modulus can be obtained in linear response theory from the shear component of the stress tensor as:
\begin{equation}
T_{xy}\,=\,\mathcal{G}\,\varepsilon\,+\,\mathcal{O}(\varepsilon^2)\,,\quad \qquad \mathcal{G}\,=\,\Re\left[G^R_{T_{xy},T_{xy}}\left(\omega=k=0\right)\right] \label{Gdef}
\end{equation}
with $G^R_{T_{xy},T_{xy}}$ the retarded Green function of the stress tensor operator. Let us notice that the shear modulus $\mathcal{G}$ pertains just the static or zero frequency response to a shear strain and it is purely non dissipative. 

The same can be done for the bulk modulus $\mathcal{K}$ from the correlator of the trace part of the $T_{ij}$, 
that is, as $\sigma^i_i  = \mathcal{K} \; \partial \cdot \phi+\mathcal{O}((\partial \cdot \phi)^2)$. 
This is compatible with the another notion of the bulk modulus, as simply  the inverse of the compressibility, which applies to more general systems, 
\begin{equation}
\mathcal{K}\,=\,-\,\mathcal{V}\,\frac{dp}{d\mathcal{V}} \label{Kdef}
\end{equation}
where $\mathcal{V}$ is the volume of the system and $p \equiv T_{xx}$ the mechanical pressure.

Another simple and important parameter in order to characterize different kind of materials is the so-called Poisson's ratio. It parametrizes how much a material compresses (or dilates) in the transverse direction when under an applied  axial tensile strain,
\begin{equation}
{\cal R}\,=\, -\,\frac{\varepsilon_{trans}}{\varepsilon_{axial}}~.
\end{equation}
It is possible to express this ratio in terms of the elastic moduli \cite{10.2307/52148}, by re-writing the elastic response as  $\varepsilon_{ij}  =\frac{1}{E} \left[ (1+{\cal R}) \sigma_{ij}- {\cal R} \sigma_{kk} \delta_{ij} \right] $, with $E$ the Young modulus. It follows that 
\begin{equation}
{\cal R}\,=\,\frac{(d-1)\mathcal{K}\,-2\mathcal{G}}{(d-1)(d-2)\mathcal{K}\,+2\mathcal{G}}  ~. \label{poisson}
\end{equation}
This allows us to classify models accordingly to their linear elastic properties. By construction, in two spatial dimension the Poisson ratio is bounded $-1<{\cal R}<\frac{1}{d-2}$, and the most auxetic behaviour corresponds to ${\cal K}\ll {\cal G}$.

Notice that ${\cal R}$ can be negative, giving a rather exotic type of response where the material actually dilates in the transverse directions. Materials of this type are called {\it auxetic} and have a number of applications. Let us advance one of  the results of Section \ref{sec:CFT} is that we will construct planar black hole solutions that are auxetic in exactly the same sense as this.

As we review  in Section \ref{sec:EFT}, the elastic moduli determine completely the speed of propagation of transverse and longitudinal phonons in homogeneous and isotropic solids.

\subsection{Phonons and solid EFTs}
\label{sec:EFT}

Let us now review how the two elastic moduli determine  the speed of propagation of transverse and longitudinal phonons, $c_{T,L}$, with no additional microscopic information on the  solid characteristics required. The relations are long known but we find illustrative to derive them using  Effective Field Theory  (EFT) methods, by treating the phonons as the Goldstone bosons associated to the spontaneous symmetry breaking pattern which takes place in solids.
The resulting {\it Solid EFT} are discussed at lowest order in derivatives for the phonon fields in \cite{Leutwyler:1996er,Dubovsky:2005xd,Nicolis:2015sra,Nicolis:2013lma}, see also \cite{Alberte:2018doe}. 
Let us emphasize that the EFT description below corresponds to the spontaneous breaking of spacetime symmetries.

We want to work with  dynamical degrees of freedom that are in a sense responsible for the spontaneous breaking of the translations that take place in solids. 
In fact in the language of the previous subsection, the spontaneous breaking of the translations can be ascribed to 
the scalar fields $\Phi^I$, taking the vacuum expectation value (vev) $\Phi^i=x^i$ in equilibrium. 
Looking at \eqref{phiphi}, then one identifies the phonons as $\phi^i$,  the perturbations around this vev (or `background').

The EFT can be formalized more sharply by labeling the set  of scalar fields with an `internal' index, so from now on we swictch to the notation $\Phi^I(x)$.
In $d$ spacetime dimensions, we need  $d-1$ scalars, so the internal index runs over $I=1,\dots,d-1$.
The theory can be then viewed as having an internal symmetry group given by the two-dimensional Euclidean group, $ISO(d-1)$, acting on $\Phi^I$ like standard translations and rotations in the internal space. 
The equilibrium configuration of an homogeneous and isotropic material is identified with the vev
\begin{equation}\label{equil}
{\Phi}^I_{eq}  =  \delta^I_j \, x^j\,.
\end{equation}
which breaks the symmetry group $ISO(d-1)\times ISO(d-1,1)$ down to the diagonal subgroup preserved by  \eqref{equil}.

At lowest order in derivatives, the effective Lagrangian is a free function of the scalar field strength 
\be\label{IIJ}
\mathcal I^{IJ}= g^{\mu\nu}\partial_\mu\Phi^I\partial_\nu\Phi^J
\ee
traced with the internal $ISO(d-1)$ symmetry group metric $\delta_{IJ}$. 
We denote by  $g_{\mu\nu}$ the spacetime metric, which it is assumed to be Minkowski.
Since $\mathcal I^{IJ}$ is a  rank $d-1$ matrix, there are $d-1$ invariants that can be split into $Z=\det \big(\mathcal I^{IJ} \big)$, and $d-2$ traces $X_n=\mathrm{tr }\, \big\{\mathcal (I^{IJ})^n \big\}$. 
Then, the most general effective action at lowest order in derivatives can be written as \cite{Leutwyler:1996er,Dubovsky:2005xd,Nicolis:2015sra,Nicolis:2013lma}
\begin{equation}\label{EFTaction}
S= -\int d^dx\,\sqrt{-g}\, V\big(Z,\{X_n\}\big)\,.
\end{equation}
The function $V(Z,\{X_n\})$ is arbitrary (up to stability constraints) and its form depends on the solid material in question.

The phonons fields are identified as the small excitations around the equilibrium configuration,  
$\phi^I= \Phi^I - {\Phi}^I_{eq}$.

In order to see that the effective action \eqref{EFTaction} encodes a mechanical response, it is illustrative to show how the elasic moduli are determined by the form of $V$. One can find these moduli by introducing a small strain  as the configuration
$$
\Phi^I  = \left( \delta^I_j  + \varepsilon^I_j \right)\, x^j
$$
with a small matrix $\varepsilon^I_j$ and working how the stress tensor depends at linear order on $\varepsilon^I_j$. The details of the computation are deferred to Appendix~\ref{app}. Let us note, however, that the expressions simplify remarkably once one uses the variable
$x_n\equiv X_n / Z^{n/(d-1)}$ instead of $X_n$ -- that is we view $V$ to be a generic function of $Z$ and $x_n$. For the bulk modulus we obtain 
\begin{equation}
\mathcal{K} = 4 Z^2 V_{ZZ} + 2 Z V_Z %= 4 \,Z^{3/2} \,\partial_Z \left( \sqrt{Z}\, V_Z \right)
\end{equation}
and for the shear modulus
\begin{equation}
\mathcal{G} = 2 \sum_{n=1}^{d-2} n^2 \frac{\partial V}{\partial x_n}~.
\end{equation}

The energy density and pressure for the background configuration read
$$
\rho=V \quad\quad {\rm and} \quad\quad p= 2\, Z\, V_Z  - V\,,
$$
and as a result the standard link between the sound speeds and the moduli follows,
\begin{equation}\label{speeds-0}
c_T^2\,=\,\frac{\mathcal{G}}{\rho+p}\,c^2,\qquad \; \; c_L^2\,=\,\frac{\mathcal{K}\,+\, 2\frac{d-2}{d-1} \mathcal{G}}{\rho+p} \, c^2~.
\end{equation}
Note that in the non-relativistic limit where the mass density $\rho_m$ dominates so that $\rho+p \simeq \rho_m c^2 +{\cal O}(c^0)$, and  one recovers the classic expressions \cite{landau7,Lubensky}  
\begin{equation}\label{speeds-02}
%c_T^2\,=\,\frac{\mathcal{G}}{\rho+p}\,c^2,\qquad \; \; c_L^2\,=\,\frac{\mathcal{K}\,+\, 2\,\frac{d-2}{d-1} \mathcal{G}}{\rho+p} \, c^2~.
c_T^2\,\simeq\,\frac{\mathcal{G}}{\rho_m} \,, \qquad \;\;  c_L^2\,\simeq\frac{\mathcal{K}\,+2\,\frac{d-2}{d-1}\,\mathcal{G}}{\rho_m}
\end{equation}
which do not involve the speed of light.\\[-2mm]

Equation \eqref{speeds-0}, together with the definition of the bulk modulus \eqref{Kdef}, implies a general relation between the longitudinal and transverse sound speeds,
\begin{equation}\label{cs-general}
\boxed{c_L^2\,= \, \frac{dp}{d\rho}\Big|_{{}_\Box} \;c^2\,+\, 2\,\frac{d-2}{d-1} \,  c_T^2}
\end{equation}
where the $|_{{}_\Box}$ subscript here is to remind that the derivative is taken while keeping zero shear strain, that is, with no shape deformation. In particular in the {\it fluid limit} (${\cal G}\to0$), this expression recovers the usual relation between the sound speed and the equation of state in fluids.

The relation between the sound speeds \eqref{speeds-0} supersedes the one that was found  for the particular case of a conformal solid in \cite{Esposito:2017qpj}, 
\be\label{SumRule}
c_L^2\,=\,\frac{1}{d-1}\,c^2\,+\,2\,\frac{d-2}{d-1}\,c_T^2~.
\ee
In Sec.~\ref{sec:SIS} we will see how the formula \eqref{speeds-0} (valid for a general Solid EFT) simplifies when one imposes Scale Invariance, and how it reduces to the one found in \cite{Esposito:2017qpj} for the conformal sub-case.

Lastly, we remark that the Poisson ratio \eqref{poisson} depends only on ${\cal G}/ {\cal K}$ and therefore it can be expressed entirely in function of the ratio of the two speeds, $c_T/c_L$. Using  \eqref{speeds-0}, one finds
\begin{equation}
{\cal R}\,=\frac{1-2\,\frac{c_T^2}{c_L^2} }{d-2-(d-3)\,\frac{c_T^2}{c_L^2}}.
\end{equation}
The auxetic limit (${\cal R}=-1$) corresponds transverse sound as close as possible to longitudinal sound, specifically 
$c_L^2=2\,\frac{d-2}{d-1} \,c_T^2$ (which follows also by setting ${\cal K}=0$ in \eqref{speeds-0}).

\section{Solids with spontaneously broken scale invariance}
\label{sec:SIS}

Spontaneously broken Scale Invariance (SI) is easy to be implemented in the solid EFT framework. We simply require the Lagrangian to be invariant under scale transformations,  which however are not a symmetry of the ground state. 
Physically, scale transformations are just a rescaling of the coordinates. 
However, in this theory there are two objects that play the role of coordinates: the `external' coordinates  $x^\mu $ and the internal space coordinates (solid element positions) $\phi^I$. Therefore, it is  conceivable that a rescaling of coordinates acts (perhaps differently)  on each of them. This leads us to consider the scale transformation as
\begin{align}\label{transf}
x^\mu &\to \;\lambda^{-1} \;x^\mu \cr
\Phi^I &\to \;\lambda^{\Delta} \;\Phi^I
\end{align}
with some `weight' $\Delta$ for the fields $\phi^I$. 

To proceed, we construct SI combinations of $I^{IJ}$.
Out of the invariants $Z$, $X_n$, the combinations of the form $x_n\equiv X_n / Z^{n/(d-1)}$ are manifestly SI (for arbitrary values of $\Delta$).
The the most general  action invariant under \eqref{transf} must have $V$ which transforms homogeneously. 
This leads to  $V(Z,\{x_m\})$ of the form
\begin{equation}\label{Vw}
V_{w}(X,Z)\, = Z^{\frac{1+w}{2}} \, F\left(\{x_n\}\right)
\end{equation}
for some constant $w$ and with an arbitrary function $F\left(\{x_n\}\right)$ of only $d-2$ arguments.

The important restriction is that there is a {\it power} of, say, $Z$ which factors out, with some exponent. 
Invariance under \eqref{transf}  uniquely fixes $w$ in terms of $\Delta$, by $(1+w)(d-1)(1+\Delta)=d$, giving

\be\label{sw}
\Delta=\frac{1-(d-1)\,w}{(d-1)(w+1)} \,,\quad {\rm or} \quad  w=\frac{1-(d-1)\,\Delta}{(d-1)(\Delta+1)}~,
\ee
which is shown in Fig.~\ref{SIsolid} for $d=4$.

Several comments are in order: 
\begin{enumerate}
\item The physical meaning of the parameter $w$ in \eqref{Vw} is readily found by computing the energy density $\rho$ and pressure $p$
for the background configuration \eqref{equil} (that is, of the solid in its state of mechanical equilibrium). 
One finds that $w$ is none  other than the equation of state parameter, 
\be\label{w}
w = \frac{ p}{\rho}~,
\ee
which in turn gives $T^\mu_\mu=(1-(d-1)w)\;\rho$.
Note  that SI fixes the equation of state to be constant for arbitrary values of the energy density/pressure, that is, that the equation of state is linear,
\be\label{w2}
p=p(\rho) = w\;\rho,
\ee
{\it to all orders}. As usual, $w$ must comply with the usual The Null Energy Condition (NEC), $w>-1$, as a necessary condition order to ensure the absence of ghosts. See below for further constraints from other consistency conditions.

\item The weight $\Delta$ introduced in  \eqref{transf} plays the same role  as the scaling dimension for $\Phi^I$, formally in the same way as for scalar operators in conformal field theories CFTs.

\item The previous point immediately  raises the question: can one apply standard logic and results from CFTs?
In fact, the fields $\Phi^I$ play the role of some scalar operators, and \eqref{Vw} is formally relativistic (it is built out of $\Phi^I$ and the Minkowski metric) so one might even wonder: does \eqref{Vw} actually define a relativistic CFT?
From our perspective, the answer to both questions is {\it no}. 
Despite appearances, the theories \eqref{Vw} are not really relativistic for an essential reason: they lack a well defined Poincar\'e invariant ground state. Indeed, the only Poincar\'e invariant configuration would be $\Phi^I = 0$ (or $\Phi^I = const$, which is equivalent by the shift symmetry). However, around this `vacuum' the kinetic terms for $\Phi^I$ are not even analytic, so the theory doesn't admit a well defined vacuum. Only around configurations with nonzero gradients $\partial_\mu \Phi^I$ the theory can be  quantized perturbatively. 

The scale invariance of the theory implies that once $\partial_\mu \Phi^I =\alpha \delta^I_\mu $ then all the values of $\alpha$ are equivalent up to rescalings. Hence, there are only two distinct configurations in principle: $\alpha=1$, which is breaks part of the Poincar\'e group but admits a perturbative quantization; or $0$, which would be relativistic but isn't really well defined.  
In other words, there is no continuous controlable way to approach the unbroken symmetry case -- the theory is bound to describe only nonrelativistic states.
Therefore the theory isn't really relativistic, even if we used a relativistic-looking language in \eqref{transf} and \eqref{Vw}.

A more heuristic reason to see that the EFTs \eqref{Vw} are essentially non-relativistic is to realize that the fields $\Phi^I$ themselves play the role of spatial coordinates. Therefore, in some sense, the scale transformation \eqref{transf} acts on time and space differently (for $\Delta\neq0$).
This has important consequences at the level of reconciling the features that the SI theories \eqref{Vw} with standard CFT results.

\begin{figure}[t]
\begin{center}
\includegraphics[width=0.9\linewidth]{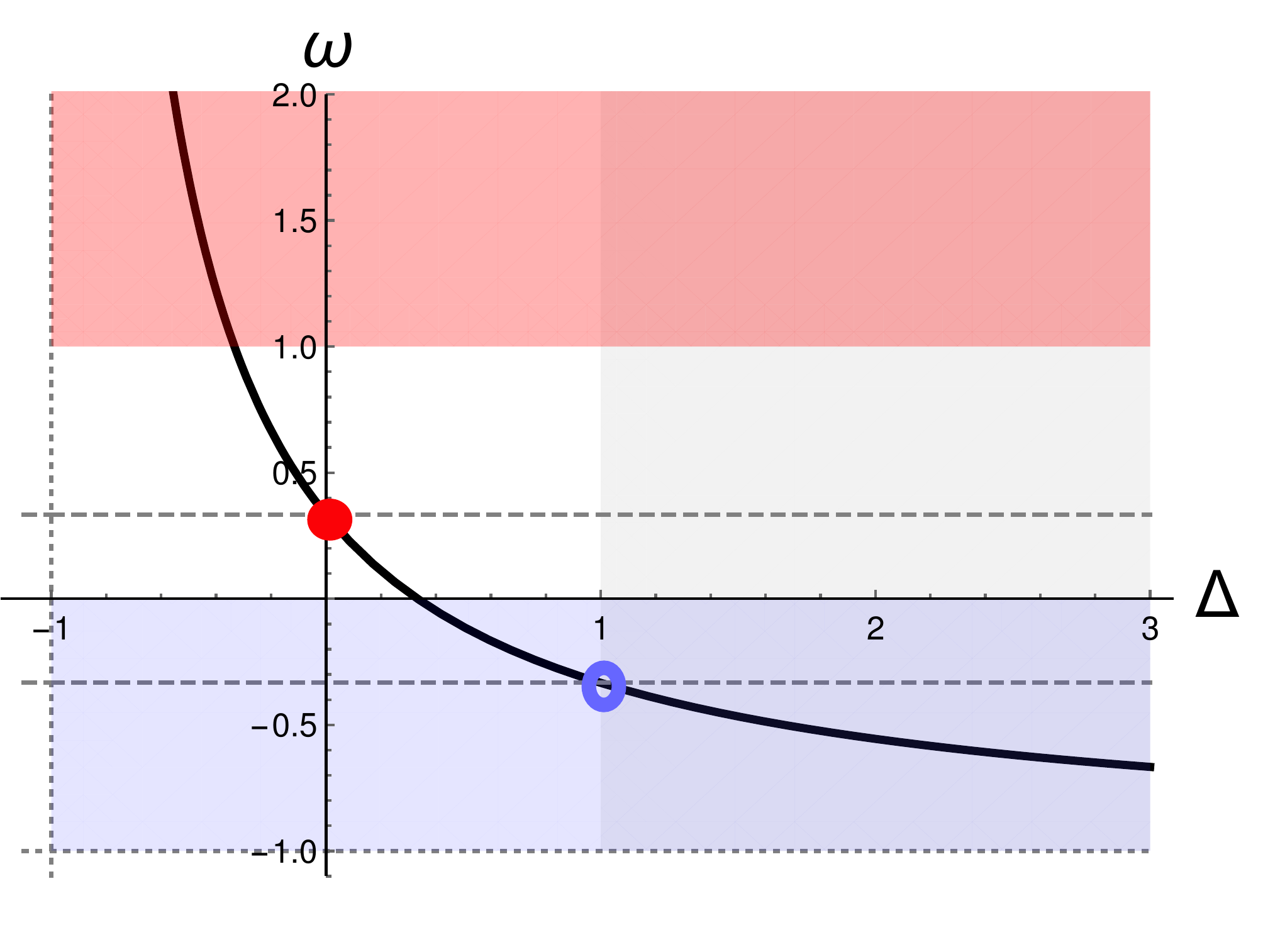}
 \caption{Relation between the equation of state parameter $w$ and the scaling dimension of the scalars $\Phi^I$ for SI solids in $3+1$ dimensions. The gray shaded area is where the naive unitarity bound $\Delta>1$ holds. 
The red shaded area ($w\geq1$) should be excluded, as the longitudinal sound speed $c_L$ is superluminal.
The red disk corresponds to the Weyl-invariant case, with $T^\mu_\mu=0$ and $\Delta=0$. 
The blue circle includes the  free theory, which is known to admit an improved the stress tensor that is traceless too.
The blue shaded region suffers from longitudinal gradient instability ($c_L^2<0$) in the limit $c_T\ll c$.}
 \label{SIsolid}
\end{center}
\end{figure}

\item The $\Delta=0$ case is special: it is invariant under local  {\it Weyl} (or {\it conformal}) transformations of the metric 
\be\label{weyl}
g_{\mu\nu} \to \Lambda^2(x^\rho)\;g_{\mu\nu}
\ee
with an arbitrary function $\Lambda(x^\rho)$ and no action on the scalars.
The symmetry group is bigger than just global SI because $\Lambda(x)$ is a free function, and this ensures vanishing trace of the stress tensor $T_\mu^\mu=0$.
This is the `conformal solid' EFT considered in \cite{Esposito:2017qpj}, which leads to the relativistic sound speeds \eqref{SumRule} and therefore doesn't seem appropriate for real world ({\it i.e.}, with nonrelativistic sound speeds) SI solids. This point is marked in Fig.~\ref{SIsolid} with a red dot. 

Let us also remark that in the examples of Sec.~\ref{sec:CFT} of solids with manifest SI the scaling dimension of the operators that are naturally identified as the  phonons also have $\Delta=0$.

\item The case $\Delta=\frac{d-2}{2}$ (corresponding to $w=-\frac{d-3}{d-1}$) with $F(\{x_n\})=x_1$ is also special: it is the free theory, which is known to admit an improved stress tensor that is also traceless, see {\it e.g.} \cite{Jackiw:2011vz,Nakayama:2013is,Rychkov:2016iqz}. This case is marked with a blue circle in Figs.~\ref{SIsolid},~\ref{SIsuperfluid}.

\item In the generic case  ($\Delta\neq 0$ or $ \frac{d-2}{2}$) the stress tensor isn't traceless. 
It isn't obvious whether one can construct an `improved' traceless stress tensor \cite{Coleman:1970je,Callan:1970ze,Polchinski:1987dy,Rychkov:2016iqz,Nakayama:2013is}, but it seems highly unlikely in the full nonlinear theory\footnote{Even ignoring interactions, the theory consists two types of `free' scalars fields, the longitudinal and transverse modes, with generically different sound speeds. In this case, one can see that the standard improvement method \cite{Jackiw:2011vz} eliminates only the contribution from one of the scalars to the trace. This already hints that there is no possible improvement and the trace is generically nonzero.}. See \cite{Nakayama:2013is,Pajer:2018egx} for discussions on this point in similar theories. 
Therefore these cases are scale but not conformal invariant ({\it i.e.} with $T^\mu_\mu\neq0$). 
One might find this surprising, however let us remind that the theories \eqref{Vw} are intrinsically nonrelativistic (see point 3) because they can be quantized only around Lorentz-breaking backgrounds.

The scale but non-conformal invariant {\it elasticity} theories \eqref{Vw} might remind the reader the well known previous example given in \cite{Riva:2005gd}. 
Despite naive similarities, however, the constructions are very different.
For instance, the example in \cite{Riva:2005gd} is a theory in Euclidean signature, and it manages to be scale- but not conformal-invariant because it lacks reflection-positivity \cite{Riva:2005gd} (see  \cite{Jackiw:2011vz,Nakayama:2013is,Rychkov:2016iqz} for reviews on this point). The SI solid EFTs instead escape conformal invariance by breaking Lorentz invariance. 

\item An intriguing feature of the relation \eqref{sw} is that the range compatible with the NEC and with gradient (in)stability,
corresponds to a surprising range in $\Delta$ that includes even negative values, as seen in Fig.~\ref{SIsolid}. It is inevitable to compare with true CFTs, where the scaling dimension of scalar operator must obey the so-called unitarity bound $\Delta>\frac{d-2}{2}$. Even the conformal solid case ($\Delta=0$, $w=1/(d-1)$) evades this bound. Again, in our view this is not a signal of violating unitarity, but simply the consequence that without full Lorentz invariance the  unitarity bound is expected to be more permissive.

\item The natural question, then, is what are the bounds on $\Delta$ that apply for the theories \eqref{Vw}. 
Stability and consistency arguments give rise to 3 types of bounds:
i) absence of ghosts (which amounts to the the NEC, $w>-1$) and which translates into  $\Delta>-1$; 
ii) absence of gradient instabilities and iii) subluminality $c_{L,\,T}<c$. We postpone this discussion to Sec.~\ref{sec:speeds}, 
once the values for the sound speeds are presented.

\end{enumerate}

\subsection{Sound speeds}
\label{sec:speeds}

Let us now return to the main point -- how SI constrains the phonon speeds.
As mentioned in point 1), SI demands that the equation of state is linear $p=w\,\rho$ to all orders in $\rho$.
In particular, this implies that the linear bulk modulus is fixed in terms of the background pressure as\footnote{This also restricts the nonlinear response for bulk strain deformations. We postpone this discussion to a forthcoming work \cite{next}.}
\be\label{KEFT}
{\cal K}=(1+w) p = w\,(\rho+p)~.
\ee
The general  formulas for the sound speeds in any (SI or not) solid EFT are
\begin{equation}\label{cLEFT}
c_T^2\,=\,\frac{\mathcal{G}}{\rho+p}\,c^2,\qquad \; \; c_L^2\,=\,\frac{\mathcal{K}\,+\, 2\frac{d-2}{d-1} \mathcal{G}}{\rho+p} \, c^2~.
\end{equation}
Plugging \eqref{KEFT} into those, one obtains
\be\label{SRw}
\boxed{~c_L^2\,=w\,c^2\,+\,2\,\frac{d-2}{d-1}\,c_T^2~.}
\ee
The first evident remark is that, once $w\ll1$, this equation allows that the two sound speeds are small, while the solid being SI. 

Next, we discuss the stability/constitency constraints. Absence of gradient instability and subluminality in the transverse sector only places a constraint on the shear modulus
\be\label{Gconstr}
0<\mathcal{G} \leq \rho+p
\ee

\begin{figure}[t]
\begin{center}
\includegraphics[width=0.9\linewidth]{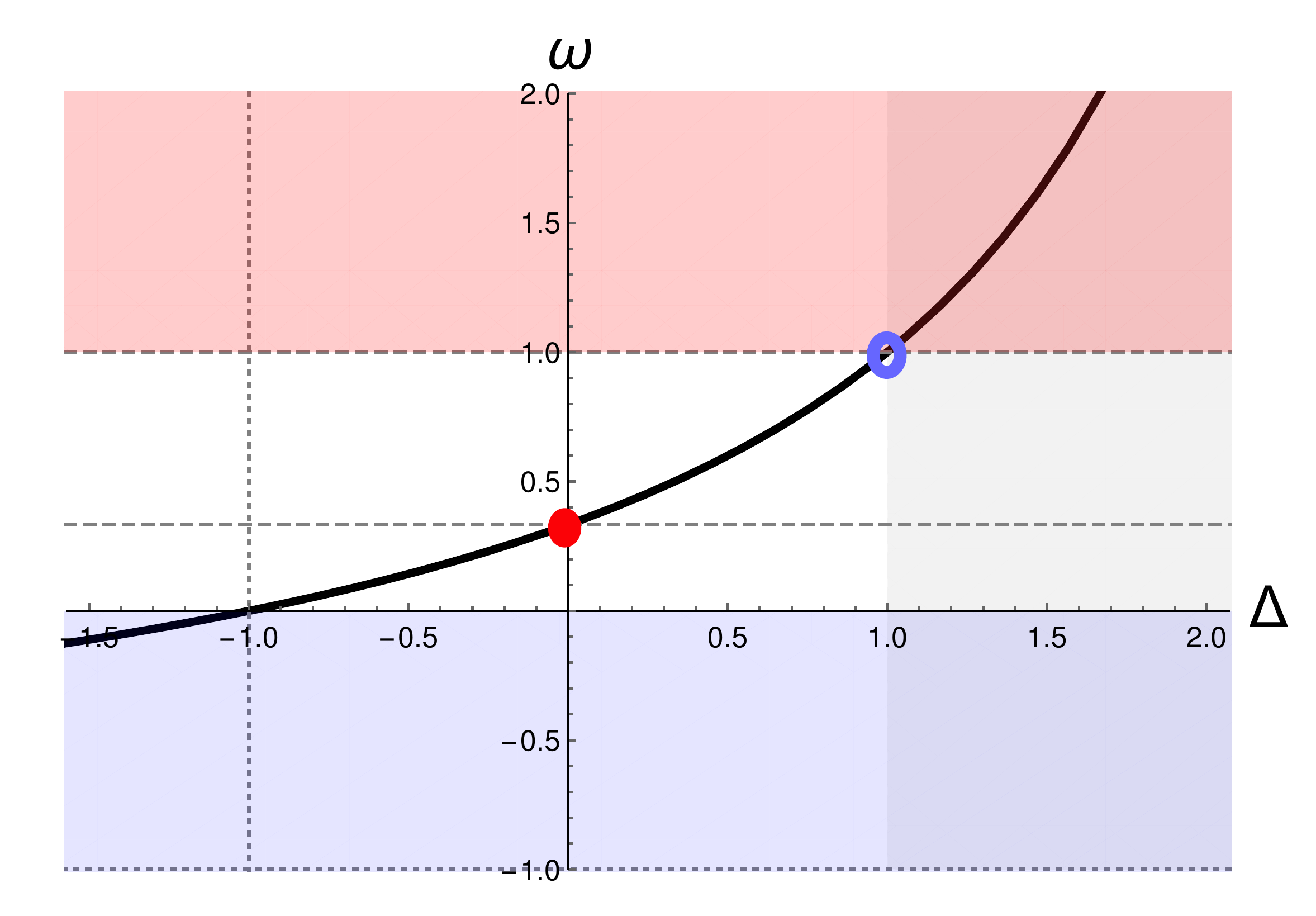}
 \caption{
The same as in Fig.~\ref{SIsolid} for a SI superfluid. }
 \label{SIsuperfluid}
\end{center}
\end{figure}

Once this is ensured, the analogous bounds on $c_L$ then constrain $w$.
The absence of gradient instability, $c_L^2>0$, places a stronger constraint on $w$ than the NEC. From  \eqref{SRw}, we find\footnote{In the exceptional case of the free theory (with $V=X$ and $c_{L\,,T}=c$), leads to $w=-\frac{d-3}{d-1}$  so \eqref{gradinst} is automatically satisfied for any $d$.}
\be\label{gradinst}
w >  - \,2\,\frac{d-2}{d-1}\, \frac{c_T^2}{c^2}~.
\ee
Since in most solids $c_{T}/ c \sim 10^{-4}$ at most, we represent this constraint in Figs.~\ref{SIsolid} and \ref{SIsuperfluid} as basically excluding the region $w<0$.
From \eqref{sw}, this translates into an {\it upper} bound on the scaling dimension $\Delta$ for SI solids $\Delta < 1/(d-1)$.

On the other hand, the subluminality condition on \eqref{SRw} requires that $w<1$, which translates into $\Delta > -\frac{d-2}{2(d-1)}$. All in all, then, we are left with allowed scaling dimensions in the window
\be\label{GIboundSolid}
\boxed{-\frac{d-2}{2(d-1)} < \Delta < \frac{1}{d-1}~.}
\ee

Finally, using \eqref{SRw} into \eqref{poisson}, one can relate the Poisson ratio in a SI solid in terms of one sound speed and $w$,
\begin{align}\label{Rw}
&\,\,\,\, \frac{c_L^2}{c^2} = \frac{(d-1)(1-(d-3) \,{\cal R} )}{{\cal R}+1}\;w \qquad\; \\&{\rm or} \qquad\; 
{\cal R} = \frac{(d-1)\,w-\frac{c_L^2}{c^2} }{ (d-3)(d-1)\,w+\frac{c_L^2}{c^2} }~.
\end{align}
Notice that  small $w$ and $c_L$ ({\it i.e.} more non-relativistic the solid) allows for more auxetic behaviour -- more negative ${\cal R}$ is allowed. Conversely, in the {\it conformal solid} limit ($w=1/(d-1)$), one finds at most ${\cal R}>0$ (from $c_L < c$).

\subsection{Scale invariant superfluids}

It is easy to extend the the above analysis of solids that realize SI to a simpler case, namely a {\it SI relativistic superfluid}, which consists in a single scalar field that has a temporal vev for the gradient, $\Psi = t + \psi$. This case has also been studied in \cite{Pajer:2018egx}.
The most general action at leading order in derivatives is $S=\int d^dx P(X_{{}_{(\Psi)}})$ where $X_{{}_{(\Psi)}}\equiv\partial_\mu\Psi g^{\mu\nu}\partial_\nu\Psi$. Scale invariance also allows for 
a nontrivial scaling dimension for $\Psi$, $\Delta_{{}_{(\Psi)}}$,  defined similarly to \eqref{transf} and it
restricts the Lagrangian to be power law in $X_{{}_{(\Psi)}}$. The only difference with respect to the SI solid case is how the exponent relates to  $w$.  
One can easily find that  $P_w(X_{{}_{(\Psi)}})=X_{{}_{(\Psi)}}^{\frac{w+1}{2}}$, and thus the $\Delta-w$ relation is now
\be\label{swSF}
\Delta=\frac{(d-1)w-1}{w+1} \,,\quad {\rm or} \quad  w=\frac{1-(d-1)\Delta}{(d-1)(\Delta+1)}~,
\ee
which is shown in Fig.~\ref{SIsuperfluid} for $d=4$. Intrerestingly, the Weyl/conformal-invariant case also requires $\Delta=0$. In this case, there is only one speed of sound $c_s^2=w \,c^2$ so the constraints from (no) gradient instability and subluminality are $0<w<1$, which gives 
%$\Delta\leq \frac{d-2}{2}$
\be\label{GIboundSF}
-1 < \Delta < \frac{d-2}{2} ~ \qquad\;\; {\rm -~for~SI~superfluids} 
\ee

\section{Solids with manifest scale invariance}
\label{sec:CFT}

Let us focus now on the other possibility mentioned in the introduction: that scale invariance (SI) is in fact manifest in the low energy theory, namely as  a nontrivial infrared fixed point of the renormalization group. 
The main physical effect of having the phonons as part of nontrivial IRFP is that the phonons are not isolated degrees of freedom, so one expects that they inevitably have diffusive behaviour. This translates in their dispersion relation as acquiring an imaginary part, $w=c_{T,L} \,k - i \Gamma(k)$. Given that the diffusive part 
$\Gamma(k)$ scales as $k^2$, at low enough energy the dispersion relation is still basically real, linear and propagating\footnote{More precisely, the phonon propagates until the so-called Ioffe-Regel crossover, whose momentum depends on how strong is dissipation (the diffusion constant) compared to speed of propagation -- the elastic modulus. For more details see \cite{Ammon:2019wci} and in particular fig.4 therein.}. Under this condition, it is justified to focus mainly on the real part of the dispersion relation ({\it i.e.} on their the speeds), as we shall assume henceforth\footnote{The complete study of the low energy dynamics (including the dissipative terms) can be obtained using Hydrodynamic techniques. See \cite{PhysRevA.6.2401,Delacretaz:2017zxd,Ammon:2019apj}.}. 

Our goal is to study this case using holography, that is, modelling the CFT as an effective theory in AdS space and using the AdS/CFT dictionary and keeping in mind that we need an elastic sector which allows for a well defined elastic response. 
The simplest holographic model is to consider a non-dynamical $AdS_{d+1}$ space with $d-1$ scalar fields $\Phi^I$ propagating in it (with no backreaction on the metric). In Poincare coordinates\footnote{To make the distinction with the previous section clearer, lower-case latin indices $a,\,b,\, \dots$ will refer to $d+1$ dimensional coordinates, so that schematically $x^a=\{z\,,x^\mu\}$.} $ds^2=g_{ab}dx^a dx^b=(\ell^2/z^2)(dz^2+dx_\mu dx^\mu)$, 
one can take the scalar Lagrangian as in the previous section, ${\cal V}(Z,x_n)$ with some function ${\cal V}$ and the invariants constructed from 
$\mathcal{I}^{IJ}_{AdS}=\partial_a \Phi^I \partial_b \Phi^J g^{ab}$. These theories admit solutions with spatial gradients  of the form 
$\Phi^I=\delta^I_\mu x^\mu$, which break both Lorentz and SI. For a certain type of the potential $V$, the holographic interpretation \cite{Alberte:2017oqx} of the solutions is that the breaking of both Lorentz and SI is spontaneous.

Given that the stress tensor plays a prominent role in elasticity, and that a model with non-dynamical metric is interpreted holographically as a theory with no stress tensor operator (see e.g. \cite{Rychkov:2016iqz}), we shall not discuss the limit of no backreaction below.  

Nevertheless, this simple model allows us to highlight an important point. The {\it holographic CFT} constructed as an AdS dual is meant to represent the infrared fixed point (IRFP) which controls the quantum critical material at low energies. Therefore it is clear that it only stands for the effective field theory description that in a sense {\it emerges} at low energies -- SI itself is an emergent symmetry for IRFPs. It is conceivable, then, that other symmetries might be emergent as well. 

In the present context, it is particularly relevant to include the possibility that the IRFP exhibits an {\it emergent 
Lorentz invariance}. 
Besides the fact that this allows us to treat the IRFP as a true CFT (invariant under emergent boost invariance, SI and special conformal transformations), 
what this means in practice is that the field theory is characterized by a well defined light cone speed $c_e$, generically different from the speed of light $c$. In order to comply with the fundamental principles of special relativity one needs to have  $c_e<c$ and (in fact the limit of interest will be $c_e\ll c$). 
The possibility that Lorentz invariance arises as an emergent symmetry has been studied {\it e.g.} in \cite{Chadha:1982qq,Nielsen:1978is,Vafek:2002jf,GrootNibbelink:2004za,Iengo:2009ix,Giudice:2010zb,Anber:2011xf,Pujolas:2011sk,Bednik:2013nxa}. 
For the present work, we shall only take this as an assumption in order to construct a simple model. 

The emergence of LI can be formulated a bit more precisely by saying that in addition to the standard (fundamental) Minkowski metric $\eta_{\mu\nu}$ the theory contains (or produces dynamically) a spin-2 object in addition to the standard Minkowski metric, and that the all the CFT operators couple to this emergent metric. To distinguish it from the fundamental Minkowski metric, we will denote it as $\eta^e_{\mu\nu}$. By definition, it allows to define an emergent line-cone structure
$$
ds^2_e = - c_e^2 \;dt^2 + dx^i dx^i~.
$$
In terms of the usual spacetime coordinates with homogeneous dimensionality $x^\mu = \{ c\,t , x^i\,\}$, the fundamental metric reads simply $\eta_{\mu\nu}={\rm diag}(-1,1,...,1)$, and the emergent one 
\be\label{etae}
\eta^e_{\mu\nu}={\rm diag}\left(-\frac{c_e^2}{c^2},1, ... , 1\right)~.
\ee
The statement that the CFT exhibits emergent Lorentz invariance then amounts to saying that all the correlators are constructed for instance from the emergent covariant distance $\Delta x^\mu\;\Delta x^\nu \eta_{\mu\nu}^e$.

Another clear consequence is that two different notions of {\it trace}, which is especially relevant for the  stress tensor operator $T^{\mu\nu}$.
Particularizing to the perfect fluid form $T^{\mu\nu}={\rm diag}(\rho,p,...p)$ that describes the solid in the homogeneous background, the two traces yield
\be\label{traces}
T^{\mu\nu}\eta_{\mu\nu} = - \rho + (d-1)\,p \;, 
\qquad{\rm and }\qquad 
T^{\mu\nu}\eta^e_{\mu\nu} = - \frac{c_e^2}{c^2}\rho + (d-1)\,p~.
\ee
This shows that the notion of an emergent conformal theory 
\be\label{traces2}
T^{\mu\nu}\eta^e_{\mu\nu}=0
\ee
is perfectly compatible with the usual non-relativistic limit (in the sense that $p/\rho\ll1$) which is required for real-world materials, so long as $c_e\ll c$. 

Another important lesson from \eqref{traces} is that the emergent light cone speed is related to the equation of state parameter ($w=p/\rho$) of the background, 
\be\label{cew}
c_e = \sqrt{(d-1)\,w}\,c~.
\ee
Snooping for a moment at Eq. \eqref{SumRule2} this will match with Eq. \eqref{SRw}  found in Section 3 for the spontaneous breaking case. 
Let us remark, however, that the symmetries are very different in the two cases.

In retrospect, assuming that the IRFP has emergent Lorentz invariance with $c_e\ll c$ use the standard  AdS/CFT dictionary in a setup that is  non-relativistic in the sense that the speeds are small compared to the speed of light. The only conceptually  important point is that the light-cone structure of the AdS space is also characterized by the emergent speed $c_e$. Since it is convenient to use natural units where the speed parameter doesn't appear explicitly, the only point to keep in mind is that natural units are those where $c_e=1$. Alternatively, one can keep track of the factors of $c_e/c$ by simple dimensional analysis as above.

In the rest of this Section, we will consider this precise setup (we model the  IRFP as an emergent CFT with small $c_e$ light-cone speed parameter) 
and study the elastic response. In order to do this, we introduces an elastic sector as a set of scalars with nonzero spatial gradients $\partial_a \phi^I$ as before, but we allow it to couple to the  stress tensor (in order to extract the elastic response in the usual fashion).
In the AdS picture this means that the scalars backreact on the metric, which will be AdS${}_4$ only asymptotically.

This analysis follows very closely the steps of \cite{Baggioli:2014roa,Alberte:2015isw,Alberte:2017cch,Alberte:2017oqx,Alberte:2016xja,Baggioli:2018bfa,Andrade:2019zey} (see also \cite{Vegh:2013sk,Blake:2013owa,Andrade:2013gsa,Andrade:2017cnc,Baggioli:2018vfc,Baggioli:2018nnp,Grozdanov:2018ewh,Grozdanov:2018fic,Amoretti:2017frz,Amoretti:2018tzw,Esposito:2017qpj} for other treatments of the elastic response), with the only main difference that now we keep in mind that the light-cone speed is emergent and therefore we treat it basically as a new parameter.

As we shall see, in this case what happens is that transverse and longitudinal sound speeds satisfy\footnote{Strictly speaking, we show this relation below for  $d=3$ and the $d$-depence is only an `educated guess' here. However the main point in the present discussion how the emergent speed $c_e$ enters.}
\be\label{SumRule2}
\boxed{c_L^2\,=\,\frac{1}{d-1}c_e^2\,+\,2\,\frac{d-2}{d-1}\,c_T^2~.}
\ee
This is structurally the same as derived for the conformal solid EFT -- it is the same as \eqref{SumRule} but replacing $c\to c_e$.
However, the quantitative difference is huge in the $c_e\ll c$ limit. The longitudinal sound speeds is now bounded by $\frac{1}{d-1}c_e^2$, and so taking $c_e$ of order $~10^{-4}c$ brings us to realistic sound speed values.

Another interesting aspect of the holographic models such as the one presented here is that they allow to
characterize rather systematically the elastic properties of the different models. In particular we will see that these models allow for one can have a significant auxetic behaviour.
Lastly, these models also incorporate finite temperature effects in a straightforward way. These effects are important because it is possible to capture the melting transition that happens at sufficiently large temperature\footnote{To be precise, this transition is totally continuous and different from the first order typical melting phase transition -- ice to water. There are some models \cite{Baggioli:2018bfa} where this transition can be discontinuous, but second order.}. Needless to say, the mechanical response changes drastically above or below the melting crossover.

\subsection{Holographic models}

As mentioned above, we shall model the low energy 'critical' behaviour using the standard holographic dictionary that maps  CFT to gravitational physics in asymptotically  AdS spacetime. For simplicity, from now on we assume that the CFT lives in $2+1$ spacetime dimensions, so that the gravitational dual is  AdS${}_4$.

By assumption, the CFT contains operators that can be identified with the displacement vectors. Their dual incarnation in the AdS${}_4$, are an identical a set of fields,  $\Phi^I$, which propagate into the holographic dimension too. The equilibrium configuration for the scalars is $\Phi^I=x^I$
and it defines the equilibrium configuration which breaks the $2+1$ Poincar\'e group and more precisely translations and rotations. The perturbations of those scalars around equilibrium encode the mechanical deformations of the system and directly the strain tensor as:
\begin{equation}
\Phi^I\,=\,x^I\,+\,\phi^I\,,\quad \quad \varepsilon_{ij}\,=\,\frac{1}{2}\,\left(\partial_i \phi_j\,+\,\partial_j\, \phi_i\right)
\end{equation}
The above identification permits to rewrite the full elastic response in terms of the dynamics of the scalar fields $\Phi^I$. For example an external shear deformation would simply correspond to a perturbation for which $\partial_x  \phi_y \neq 0$.

We consider the generic holographic massive gravity models introduced in \cite{Baggioli:2014roa,Alberte:2015isw}, and studied in several directions in \cite{Baggioli:2015zoa,Baggioli:2015gsa,Baggioli:2015dwa,Alberte:2016xja,Alberte:2017cch,Alberte:2017oqx}. 
The models are defined as a gravitational theory with negative cosmological constant $\Lambda$. The metric is locally  Minkowskian,
$g_{\mu\nu} \sim {\rm diag}(-c_e^2/c^2,1,1,1) + ...$, with an input speed parameter $c_e$ that is unrelated to the speed of light because the 4D space is only holographic. We can thus take $c_e\ll c$ without affecting at all the consistency of the theory, and which gives the important benefit of realizing conformal material with slow sound speeds as described in hte introduction.
In the following, we will work in the units $c_e=1$ unless otherwise stated.

The model is then defined by the following  action in the 4D bulk space,
\begin{equation}
\mathcal{S}\,=\,\int d^{4}x\,\sqrt{-g}\,\left[\,\frac{R}{2}-\Lambda-\,m^2\,V(X,Z)\,\right]\label{model}
\end{equation}
with 
$\mathcal{I}^{IJ}_{AdS}=\partial_a \Phi^I \partial_b \Phi^J g^{ab}$
%$\mathcal{X}^{IJ}=\partial_\mu \Phi^I \partial^\mu \phi^J$ 
and 
$X\equiv \frac{1}{2}\mbox{Tr}(\mathcal{I}^{IJ}_{AdS})$ and $Z=\det(\mathcal{I}^{IJ}_{AdS})$. For simplicity, we focus on $d=3$ but we will comment on generic and universal features. 

For specific choices of the potential $V(X,Z)$, the model \eqref{model} represents the gravity dual of a CFT at finite temperature and zero charge density where translational invariance is broken spontaneously. Using the standard AdS/CFT dictionary, this defines for us a CFT that will have non-zero elastic moduli and so it can be interpreted as a model for a solid in a quantum critical regime.
More precisely, a well-defined elastic response can be defined for potentials which decay at the boundary as $V \sim u^3$ or faster \cite{Alberte:2017cch}. Moreover, for potentials whose fall-off at the boundary is $V \sim u^5$ or faster, this elastic response is associated to the presence of massless propagating phonons \cite{Alberte:2017oqx}.

This field configuration admits an AdS black brane geometry
\begin{align}
&ds^2\,=\,\frac{1}{u^2}\left(-f(u)\,dt^2+\frac{du^2}{f(u)}+\,dx^i dx^j\right),\nonumber\\
& f(u)\,=\, u^3\int_u^{u_h}\,\left(\frac{3}{v^4}-\frac{m^2}{v^4}V(v^2,v^4)\,\right)d v\label{geometry}
\end{align}
We fix the cosmological constant to $\Lambda=-3$. 
 We assume the presence of an event horizon at $u=u_h$ defined by $f(u_h)=0$. The associated entropy density is $s=2 \pi /u_h^2$ and the corresponding temperature reads $T=-\frac{f'(u_h)}{4\,\pi}$.\\

The shear elastic modulus for these models can be obtained solving numerically the equation:
\begin{align}
h''\,+\,\left(\frac{f'}{f}-\frac{2}{u}\right) h'\,-\frac{2 \,m^2\, V_X(u^2,u^4)}{f}\,h\,=\,0
\end{align}
for the metric perturbation $h\equiv \delta g_{xy}$. The perturbation is assumed to be static, $\omega=0$, and the subscript $X$ indicates the derivative with respect to $X$. In order to extract the retarded correlator we have to impose ingoing boundary conditions at the horizon, see \cite{Alberte:2016xja} for more details. The UV expansion of the shear perturbation reads
\begin{equation}
h(u)\,=\,h_0\,\left(1\,+\,\dots\right)\,+\,h_3\,u^3\,+\,\dots \label{exp}
\end{equation}
where $h_0$ represents a source for the $T_{xy}$ operator and $h_3$ encodes the VEV of the stress tensor $ \langle T_{xy} \rangle$ \cite{Skenderis:2002wp}. Following equation \eqref{Gdef}, which defines the shear modulus (as used in previous works \cite{Alberte:2015isw,Alberte:2016xja,Alberte:2017cch,Alberte:2017oqx,Baggioli:2018bfa}), we simply find
\begin{equation}
\mathcal{G}\,=\,\frac{3}{2}\,\frac{h_3}{h_0}
\end{equation}
At this stage, it is important to make a stop to discuss this result. To be more precise, here we are considering the response of the stress tensor to a geometrical shear deformation $h_0$. A more physical approach is to consider the response of the stress tensor $h_3$ in terms of a mechanical shear deformation $\partial_x \phi_y \neq \phi_{s}$. At linear level, it does not make much difference. It is straightforward to check that:
\begin{equation}
\frac{h_3}{h_0}\,=\,-\,\frac{h_3}{\phi_s}
\end{equation}
The previous result can be understood noticing that the gauge invariant perturbation encoding a ``gauge invariant shear strain'' is indeed a combination of $\phi_s$ and $h_0$ \cite{Baggioli:2014roa}. Therefore it is clear the two will produce the same result. Once the framework will be extended at non-linear level, the interpretation in terms of the scalars perturbations is much more direct and simpler and it will be convenient to fix $h_0=0$ once and forever.\\
The solution for the shear modulus can be found analytically in the limit $m\ll T$ \cite{Alberte:2016xja,Alberte:2017oqx}
\begin{equation}
\mathcal{G}\,=%\,\frac{3\,\epsilon\,-\,2\,s\,T}{4}\,=\,\lim_{\mathfrak{z}\rightarrow 0}\frac{m^2}{4}\left[\frac{V_h}{u_h^3}+3 \int^{u_h}_{\mathfrak{z}} \frac{V}{\xi^4}\,d\xi\right]\label{Kmod}
m^2 \, \int_0^{u_h}\,\frac{V_X\left(\zeta^2\,,\zeta^4\right)}{\zeta^2}\,d\zeta\,+\,\mathcal{O}(m^4)\label{integral}
\end{equation}

On the other hand, the bulk modulus is defined in equation \eqref{Kdef}. Due to conformal symmetry, the stress energy tensor is traceless, thus 
\begin{equation}\label{pressure}
p=T_{ii} = T_{tt}/2\equiv \rho /2
\end{equation} 
We can guess the volume dependence of the energy density $\rho$ quite easily. 
Consider a homogeneous system with equation of state $\gamma=p/\rho$ in a box of volume 
$\mathcal{V}$. In an adiabatic process that changes volume, the energy density scales with volume as $\rho \propto \mathcal{V}^{-1-\gamma}$.
Therefore in our case 
\begin{equation}
\rho\, %\propto\, a^{-3}\, 
\propto\, \mathcal{V}^{-3/2}
\end{equation}
and the total bulk modulus is just 
\begin{equation}
\mathcal{K}\, = \, \frac{3}{4}\, \rho ~.
\label{bulkmod}
\end{equation}

\subsection{Sound speeds}

In the current setup, the phonons can be found as the poles in the $T_{ij} T_{kl}$ correlator at finite wavenumber $k$. In the gravitational dual this is done by finding the spectrum of quasi-normal modes. We shall not repeat this exercise here, since it  has already been done in  \cite{Alberte:2017cch,Alberte:2017oqx,Ammon:2019apj,Baggioli:2019abx}. The conclusion of these works is that for $b<5/2$
the spectrum of QNMs in our benchmark  of models contains gapless modes with a gapless dispersion relation  of the form  
\be\label{disp}
w=c_{s} k - i D k^2 +\dots
\ee
both for transverse \cite{Alberte:2017oqx,Alberte:2017oqx} and longitudinal \cite{Ammon:2019apj,Baggioli:2019abx} waves. 
Since the diffusive part scales as a higher power of $k$ it is still possible to preserve a clear notion of propagating sound modes and sound speeds -- at low enough $k$.

Moreover, the sound speeds of the QNMs can be found numerically by following the motion of the pole as $k$ changes \cite{Alberte:2017oqx,Ammon:2019apj,Baggioli:2019aqf,Baggioli:2019abx}. The numerical result thus obtained for the sound speed agree formally with what one expects from elasticity theory, that is  Eq.~\eqref{speeds-0},
\begin{equation}
c_T^2\,=\,\frac{\mathcal{G}}{\rho+p} 
\;,\qquad \;
c_L^2\,=\,\frac{\mathcal{G}\,+\,\mathcal{K}}{\rho+p}
\label{speeds2}
\end{equation}
with  $\rho$, $P$, $\mathcal{G}$ and $\mathcal{K}$  the energy density, pressure, and elastic the moduli respectively, for $d=3$ space-time dimensions.  
This agreement justifies the physical  identification of these modes as   physical phonons. 

The speeds in Eq.\eqref{speeds2} are expressed in the units of the light-cone speed present in the AdS theory. 
In a truly relativistic CFT (with light cone speed identical to the speed of light), the units are restored in Eq.\eqref{speeds2} trivially: by a multiplicative $c^2$ factor.

However, 
it is interesting to take the AdS theory merely as a model for a low energy CFT with emergent light-cone speed $c_e\ll c$ (in order that  sound can be slow compared to light). 
With this in mind, let us now restore the factors of  $c_e$-dependence in \eqref{speeds2}.
At the technical level, this can be done by performing a rescaling of the time coordinate, 
\begin{equation}\label{resc}
\partial_t \rightarrow \frac{c}{c_e}\,\partial_t\,,\quad \omega\,\rightarrow\,\frac{c}{c_e}\,\omega~.
\end{equation}
From this we can immediately derive\footnote{A similar conclusion can be reached by noticing that under restoring $c_e$:
\begin{equation}
\mathcal{G}\,\rightarrow \mathcal{G}\,,\quad \chi_{PP}\,\rightarrow \left(\frac{c}{c_e}\right)^2\,\chi_{PP}
\end{equation}} that:
\begin{equation}
c_{T,L}\,\rightarrow\,\frac{c_e}{c}\,c_{T,L} \label{rescale}
\end{equation} 
This has two  immediate consequences. First, it follows that the speeds satisfy
\be\label{SumRule3}
c_L^2\,=\,\frac{1}{2}c_e^2\,+\,c_T^2~,
\ee
which is the same as \eqref{SumRule2} for $d=3$. 

Second, since the rescaling \eqref{resc} affects the energy density $\rho$ but not the pressures $T_{ij}$, the full expression for the sound speeds is
\begin{equation}
c_T^2\,=\,\frac{\mathcal{G}}{(c_e/c)^2 \rho+p} \,c_e^2\,
\;,\qquad \;
c_L^2\,=\,\frac{\mathcal{G}\,+\,\mathcal{K}}{(c_e/c)^2 \rho + p} \,c_e^2~.
\label{speeds22}
\end{equation}
It is interesting to rewrite these in terms of the physical mass density, which relates to the energy density in the usual form
$$
\rho_m \equiv \rho / c^2~.
$$ 
One then finds that Eq.~\eqref{speeds22} reduces to
\be\label{cte}
c_T^2\,=\,\frac{\mathcal{G}}{\rho_m+p/c_e^2} 
\;,\qquad \;
c_L^2\,=\,\frac{\mathcal{G}\,+\,\mathcal{K}}{\rho_m+p/c_e^2} ~.
\ee
Recall that the spontaneous breaking case (Sec.\ref{sec:SIS}) leads to expressions of the form 
\be\label{ctSB}
c_T^2\,=\,\frac{\mathcal{G}}{\rho_m+p/c^2} 
\ee
where the pressure contribution in the denominator is much more suppressed. 
Therefore, one can say that the effect of having manifest SI with a `slow'  emergent cone ($c_e\ll c$) ends up enhancing the pressure contribution in the demonimator of the sound speed formulas, thereby reducing the sound speeds.

Upgrading the discussion to $d$ dimensions, and using the tracelessness condition ($p =(c_e/c)^2 \rho/(d-1)$ in $d$ dimensions) Eq.~\eqref{cte} further simplifies to
\be\label{cte2}
c_T^2\,=\, \frac{d-1}{d} \,\frac{\mathcal{G}}{\rho_m} \qquad {\rm (manifest~SI)}~.
\ee
This is to be  contrasted with the conventional expression \eqref{speeds-02}
\be\label{cte2SB}
c_T^2=\frac{\mathcal{G}}{\rho_m} \qquad {\rm (spontaneous~breaking)}
\ee
(up to tiny $O(\frac{p}{\rho_m c^2})$ corrections), 
which holds in the  EFT picture (spontaneous breaking) but is not granted to apply in the presence of an emergent CFT-like fixed point. 

Note that the difference between \eqref{ctSB} and \eqref{cte2} is independent of $c_e$. This deserves two comments. First, the discrepancy looks surprising but actually it is due to  the fact that the low energy theories are very different -- SI is realized in a completely different way in the two cases. 
Second, this implies that the discrepancy persists even in the limit $c_e \ll c$. In particular, taking $c_e/c$ in the range $10^{-5}-10^{-4}$  brings the sound speeds into the range of real-world materials so from this point of view this has a chance to correspond to a realistic material at a critical point. 
It is tempting to say that the relation betwee $c_T$, $G$ and $\rho_m$ can provide  a signature of whether a material is controlled by such peculiar IR dynamics. (A similar discrepancy arises also in the longitudinal sector.)

Let us remark that the case of spontaneous breaking of SI with emergent Lorentz symmetry at low energy represents another well defined option. A proper discussion of it is beyond the scope of this paper, but let us offer one comment. In this case, one wonders whether the EFT should be obtained from coset construction referred to the breaking of the `fundamental' Poincare group (with speed $c$) or from the emergent one (with speed $c_e$), which is also spontaneously broken in the ground state. 
If the latter option is the relevant one, then  we would expect Eq.~\eqref{cte} to apply. In this case, however, the pressure is not constrained by the emergent-tracelessness condition so one wouldn't obtain \eqref{cte2}. Still, one would also expect order-one deviations from \eqref{cte2SB}, basically due to  the enhancement of the pressure term in the denominator.

\subsection{Elastic response in a benchmark model}

After defining the linear response in abstract terms, we restrict ourselves to a specific and quite generic form of the potential $V$ to make more quantitative statements. In particular, for the rest of the paper we  consider the benchmark potential:
\begin{equation}
V(X,Z)\,=\,X^\mathfrak{a}\,Z^{\frac{\mathfrak{b}-\mathfrak{a}}{2}}\label{bench}
\end{equation}
In order to ensure the consistency of this choice \eqref{bench}, and of the model \eqref{model} in general, one must impose a number of requirements. First,  absence of ghosts, absence of gradient instabilities locally in the 4D theory leads to limit the parameters $\mathfrak{a},\mathfrak{b}$ in the range \cite{Alberte:2018doe}:
\begin{equation}
\mathfrak{a}\,\geq 0\,\land\,\mathfrak{b}\,\geq\,1
\end{equation}
Demanding the positivity of the linear elastic moduli, and of the energy density at low temperatures (see Eqs.~\eqref{ep} and \eqref{def} below) restricts $\mathfrak{b}$ further as
\begin{equation}
\mathfrak{b}\,\geq\,\frac{3}{2}~.
\end{equation}
Finally, in order to restrict to the theories where the phonons are gapless we need to further impose \cite{Alberte:2017oqx,Alberte:2017cch}:
\begin{equation}
\mathfrak{b}\,\geq\,\frac{5}{2}~.
\end{equation} 
Below $5/2$ the phonons acquire a mass gap. This can be translated as having additional (`explicit') sources that break translational invariance,  suggesting that the phonon speeds might depart from the expressions \eqref{speeds2}.

Notice that the constraints considered are purely bulk requirements and they represent just necessary but not sufficient conditions for the full consistency of our boundary field theory. In order to have a final verdict, a detailed QNMs computation would be needed.\\

\begin{figure}[t]
\begin{center}
\includegraphics[width=0.9\linewidth]{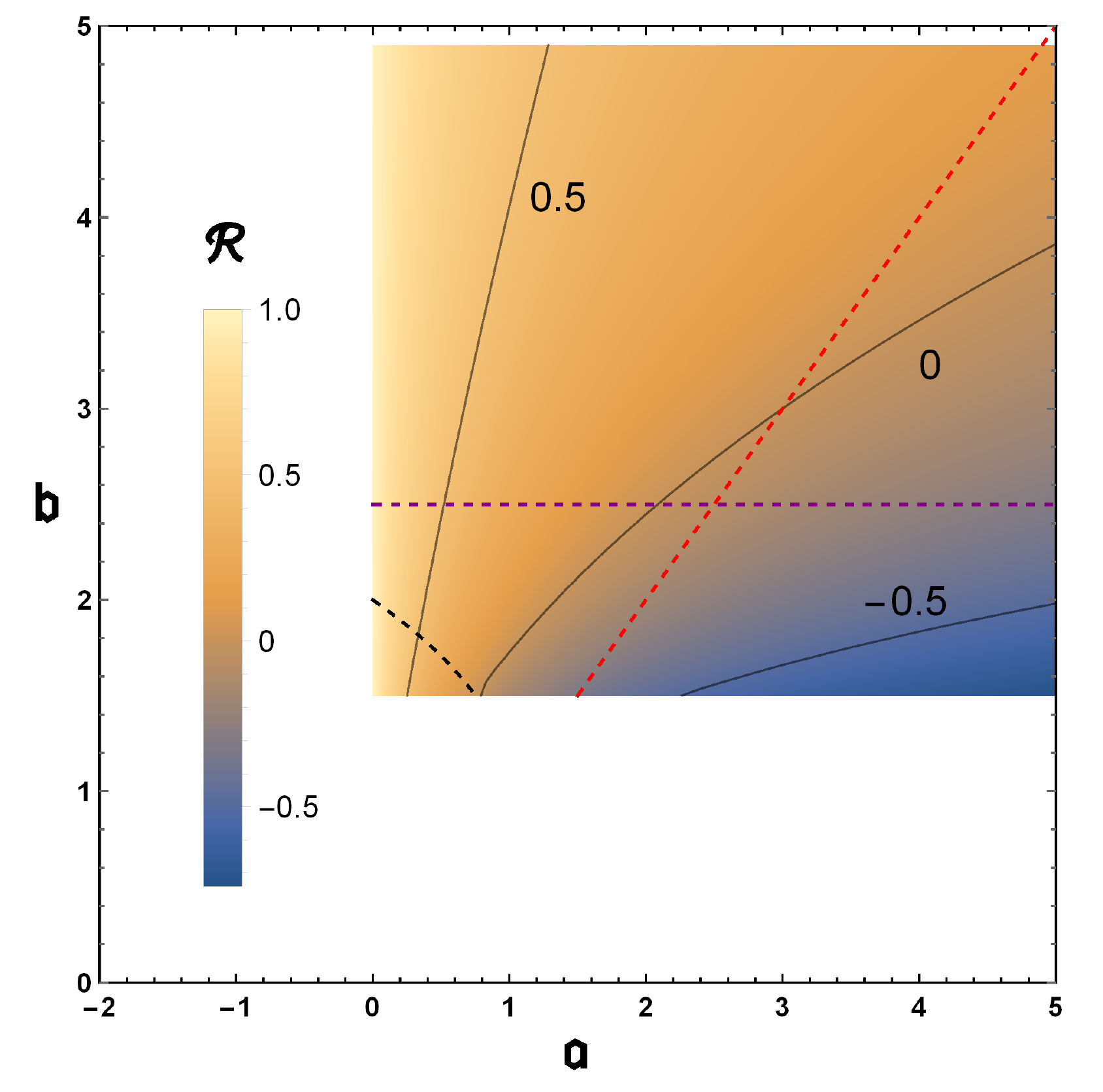}
\caption{Poisson Ratio $-1<{\cal R}<1$ for the benchmark model \eqref{bench} within the consistency region at low temperatures $T/m \ll 1$.  From orange to blue the material becomes more and more auxetic, \textit{i.e.} with a negative Poisson ratio. In the region above the purple dashed line, the phonons are gapless. The red dashed line marks the simple choice of potential $V(X,Z)=X^\mathfrak{a}$. The black dashed line shows the region where all local speeds are sub-luminal in the bulk (see Fig.~\ref{figv} for the boundary values of the phonon speeds).}
 \label{poissfig}
\end{center}
\end{figure}

Restricting ourselves to our benchmark model \eqref{bench}, we can write the expressions for the energy density and pressure of the background as
\begin{equation}
\mathcal{\rho}=\frac{1}{u_h^3}\,+\,m^2\,\frac{u_h^{2\,\mathfrak{b}\,-\,3}}{2\,\mathfrak{b}\,-\,3}
\,,\quad \quad 
p\,=\, \frac12\, \rho
\label{ep}
\end{equation}
as well as for the elastic moduli
\begin{equation}
\mathcal{G}=\frac{\mathfrak{a}}{2\mathfrak{b}-3}\,m^2\,\,u_h^{2\mathfrak{b}-3}\,+\,\mathcal{O}(m^4)\,,\quad \quad 
%\mathcal{K}\,=\,\frac{3}{4}\,\left(\frac{1}{u_h^3}\,+\,m^2\,\frac{u_h^{2\,\mathfrak{b}\,-\,3}}{2\,\mathfrak{b}\,-\,3}\right)
\mathcal{K}\,=\,\frac{3}{4}\, \rho
\label{def}
\end{equation}
and noticing immediately that the parameter $\mathfrak{a}$ is what distinguish in the static response a solid with respect to a fluid. For $\mathfrak{a}=0$, the static shear modulus is zero and the system does behave like a fluid. From  \eqref{def} we can immediately obtain the Poisson ratio ${\cal R}$ \eqref{poisson} for our holographic models. We show how ${\cal R}$ depends on $\mathfrak{a},\,\mathfrak{b}$ at low temperature in Fig.~\ref{poissfig}. \\
\begin{figure}[t]
\centering
\includegraphics[width=0.9\linewidth]{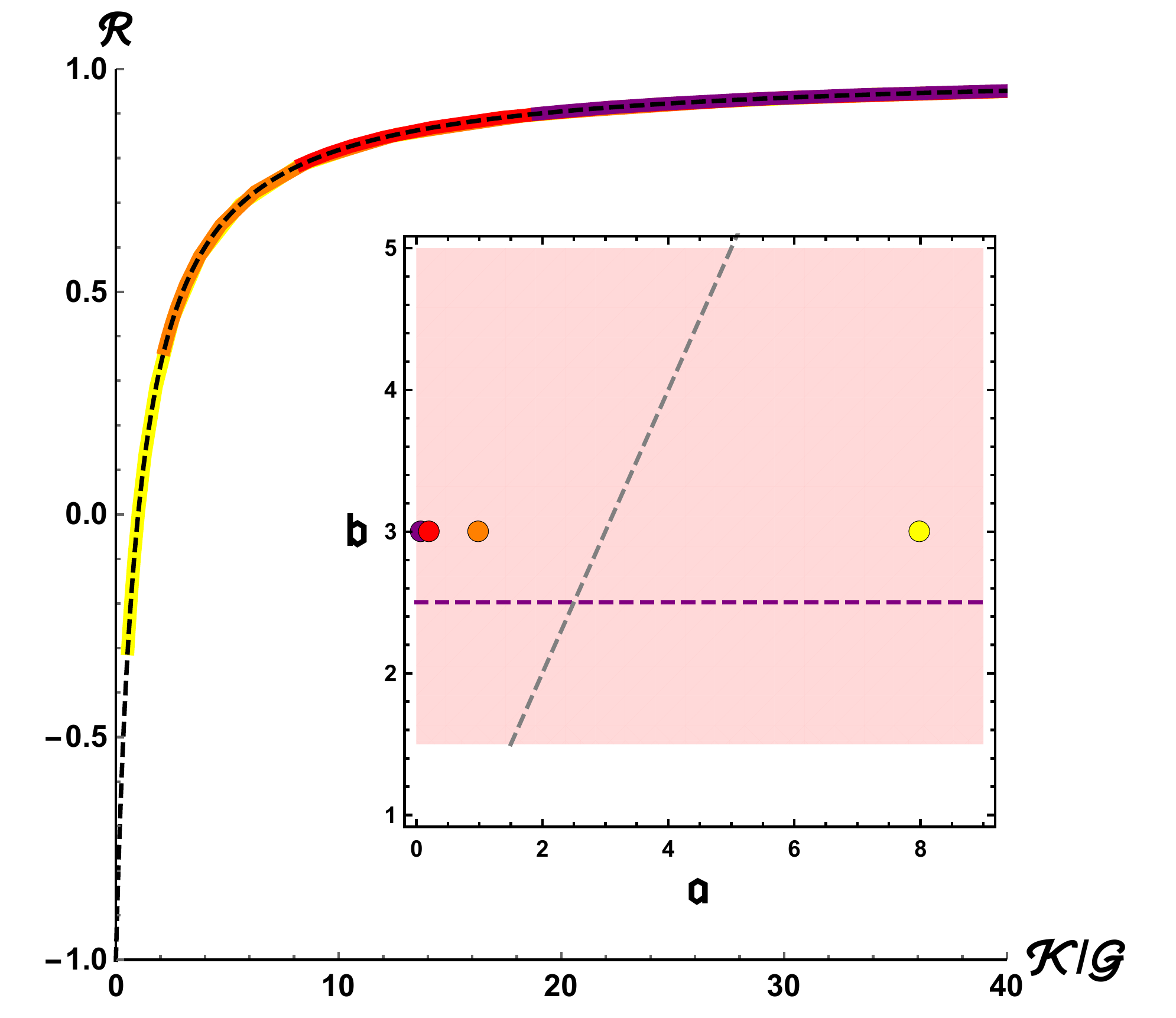}
\caption{Poisson ratio as a function of $\mathcal{K}/\mathcal{G}$ for the model \eqref{bench}. The colors represent specific points in the parameter space, which are shown in the inset, within the region of consistency. The various colored lines are produced changing the dimensionless parameter $T/m$ at fixed $\mathfrak{a},\mathfrak{b}$. }
\label{check}
\end{figure}
At large $T/m \gg 1$, the Poisson ratio always goes to the fluid value ${\cal R}=1$. This indicates that within our model, no matter the choice of the potential $V(X,Z)$, the limit of large temperature correspond to a fluid phase with a maximum Poisson ratio. We can easily understand this phenomenon by noticing that at $T/m \gg 1$ the graviton mass and therefore the additional structure induced by the scalars $\Phi^I$ are completely negligible and the phenomenology is simply the one of a relativistic strongly coupled fluid. On the contrary, at small temperature, the effects of the scalars are dominant and the phenomenology depends crucially on the choice of the potential $V(X,Z)$ and the Poisson ratio differs consistently from the fluid value. More precisely we can provide a rough classification of benchmark model \eqref{bench} as follows (see related Fig.~\ref{poissfig}):
\begin{itemize}
\item For small $\mathfrak{a}$ and large $\mathfrak{b}$ the Poisson ratio is large and close to its upper limit ${\cal R}=1$. This class of models refers therefore to incompressible and elastic materials such as rubber.
\item For $\mathfrak{a}\sim \mathfrak{b}$ the Poisson ratio is in the range $-0.5<{\cal R}<0.5$, similarly to the typical values for most steels and rigid polymers.
\item For large $\mathfrak{a}$ and small $\mathfrak{b}$ the Poisson ratio is negative (\textit{i.e.} it exhibits {\it auxetic} behaviour) and close to its lower limit ${\cal R}=-1$.  As we will see later this is correlated with the presence of superluminal speeds of sound hinting towards a possible instability.
\end{itemize}

An even better way of classifying our theories consists in plotting their Poisson artio ${\cal R}$ in function of the dimensionless quantity $\mathcal{K}/\mathcal{G}$ as for example presented in \cite{Greaves2011}. The similarities with the realistic  results are presented in Fig.~\ref{check}. As already hinted in Fig.~\ref{poissfig}, small $\mathfrak{a}$ corresponds to foam-like material whether large $\mathfrak{a}$ to rubber-like materials.\\

\begin{figure}[t]
\begin{center}
\includegraphics[width=0.9\linewidth]{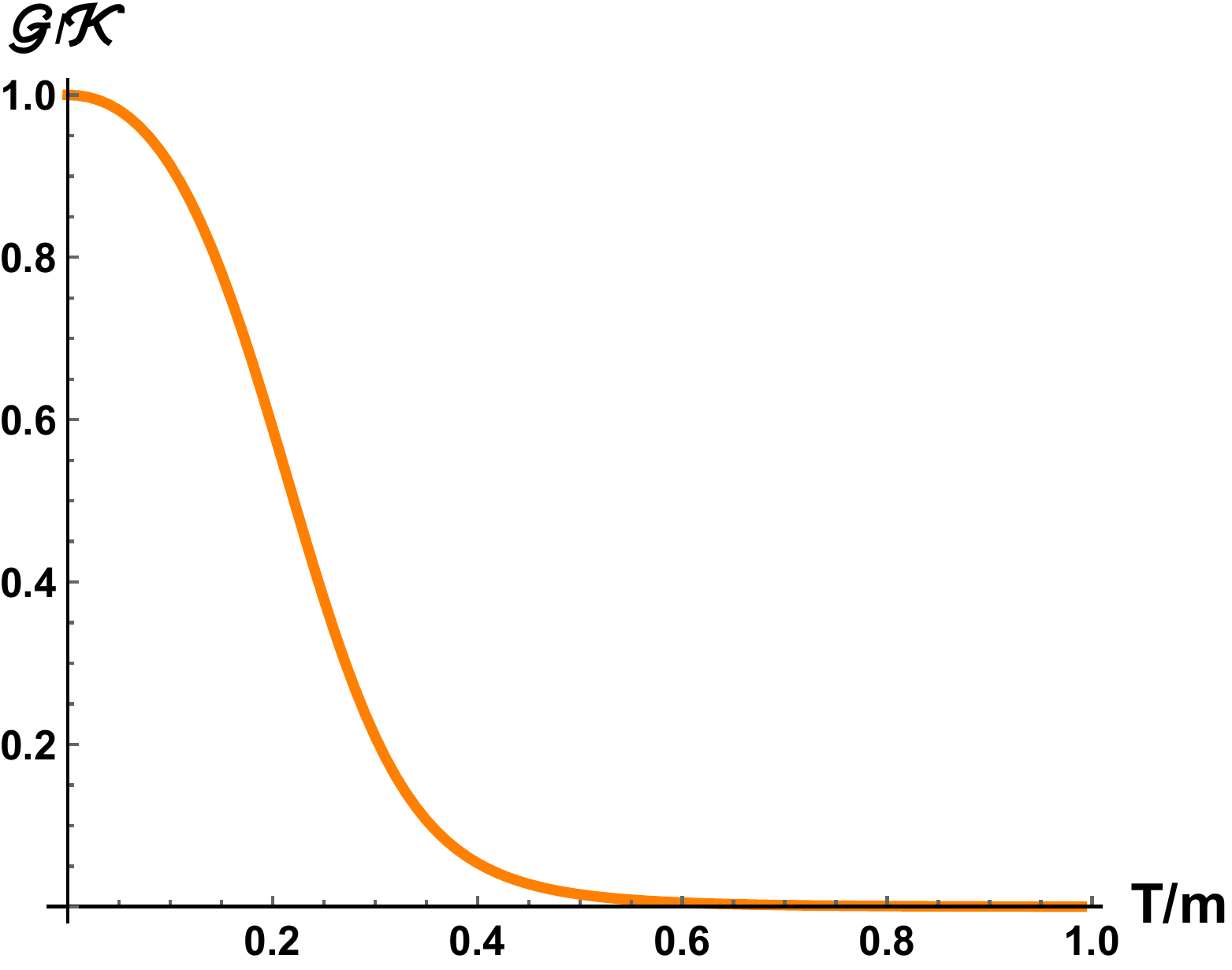}

\vspace{0.2cm}

\includegraphics[width=0.9\linewidth]{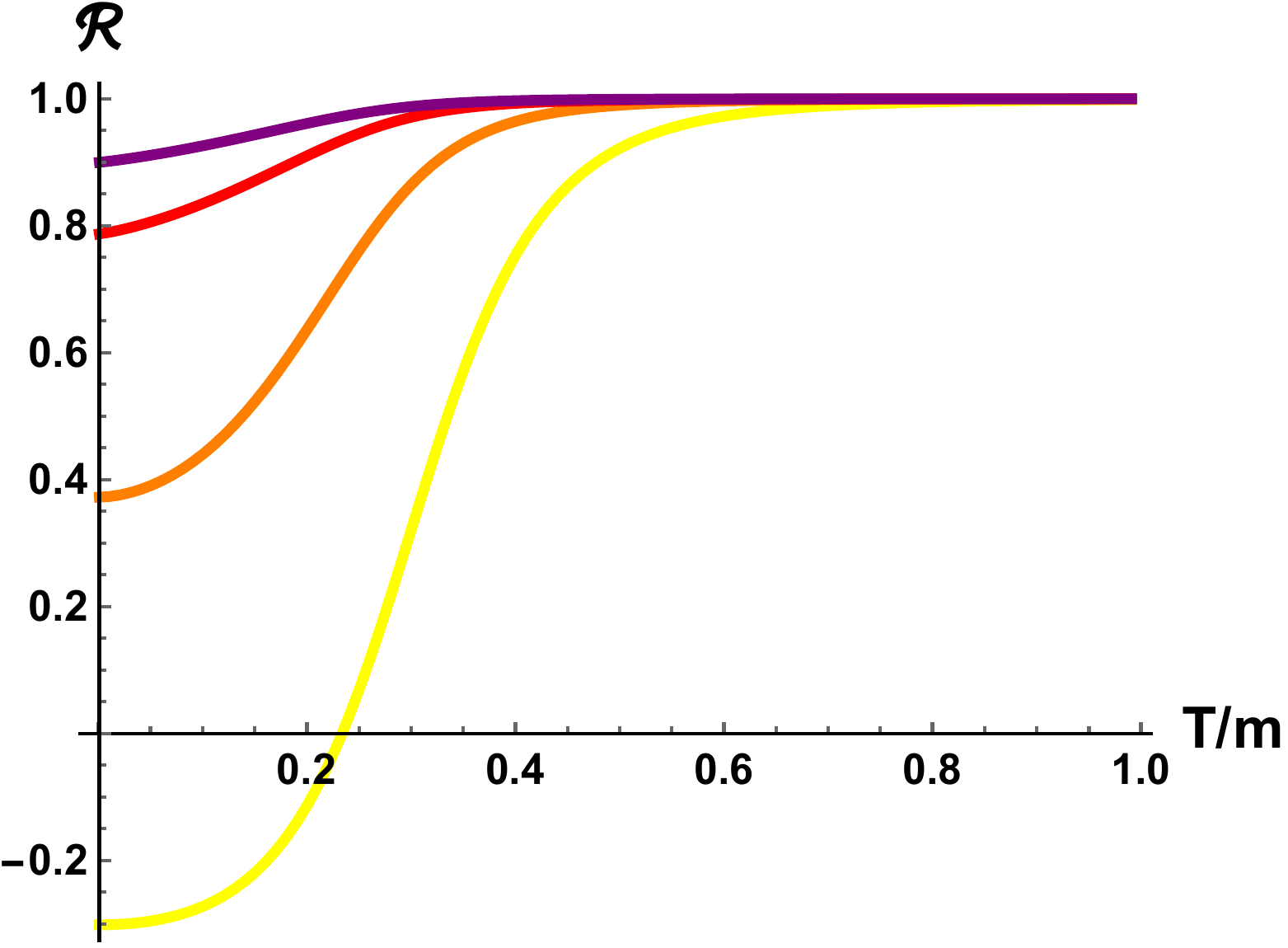}
 \caption{\textbf{Top: }The elastic moduli for the potential $V(X,Z)=X^3$ in function of $T/m$. \textbf{Bottom: }Poisson Ratio ${\cal R}$ in function of the dimensionless temperature $T/m$ for various choices of potential. The specific $(\mathfrak{a},\mathfrak{b})$ are indicated with the same colors of Fig.~\ref{check}. Notice that at large temperature the Poisson ratio always goes towards the fluid limit ${\cal R}=1$.}
 \label{linel}
\end{center}
\end{figure}

As another important feature, we can analyze the behaviour of the linear elastic moduli which are shown in Fig.~\ref{linel}. It is evident from the Figure that both the moduli  goes to zero in the limit of $T/m \gg 1$ in a continuous fashion which typical of viscoelastic and glassy materials \cite{2009PhR...476...51C}. Additionally, their ratio $\mathcal{G}/\mathcal{K}$ goes to zero at large temperatures indicating again that at $T/m \gg 1$ we are always in a fluid phase. Moreover, we can compute the behaviour of the Poisson ratio in function of the temperature (Fig.~\ref{linel}). Our results suggest that as a generic property such a ratio decreases with increasing the dimensionless ratio $T/m$. We can think of the previous properties as the ``melting'' in our holographic system, which is very similar indeed to the phenomenology of amorphous solids and glasses as already hinted in previous literature \cite{Alberte:2017oqx,Baggioli:2018qwu,baggioli2018soft}.

\begin{figure}[t]
\begin{center}
\includegraphics[width=0.8\linewidth]{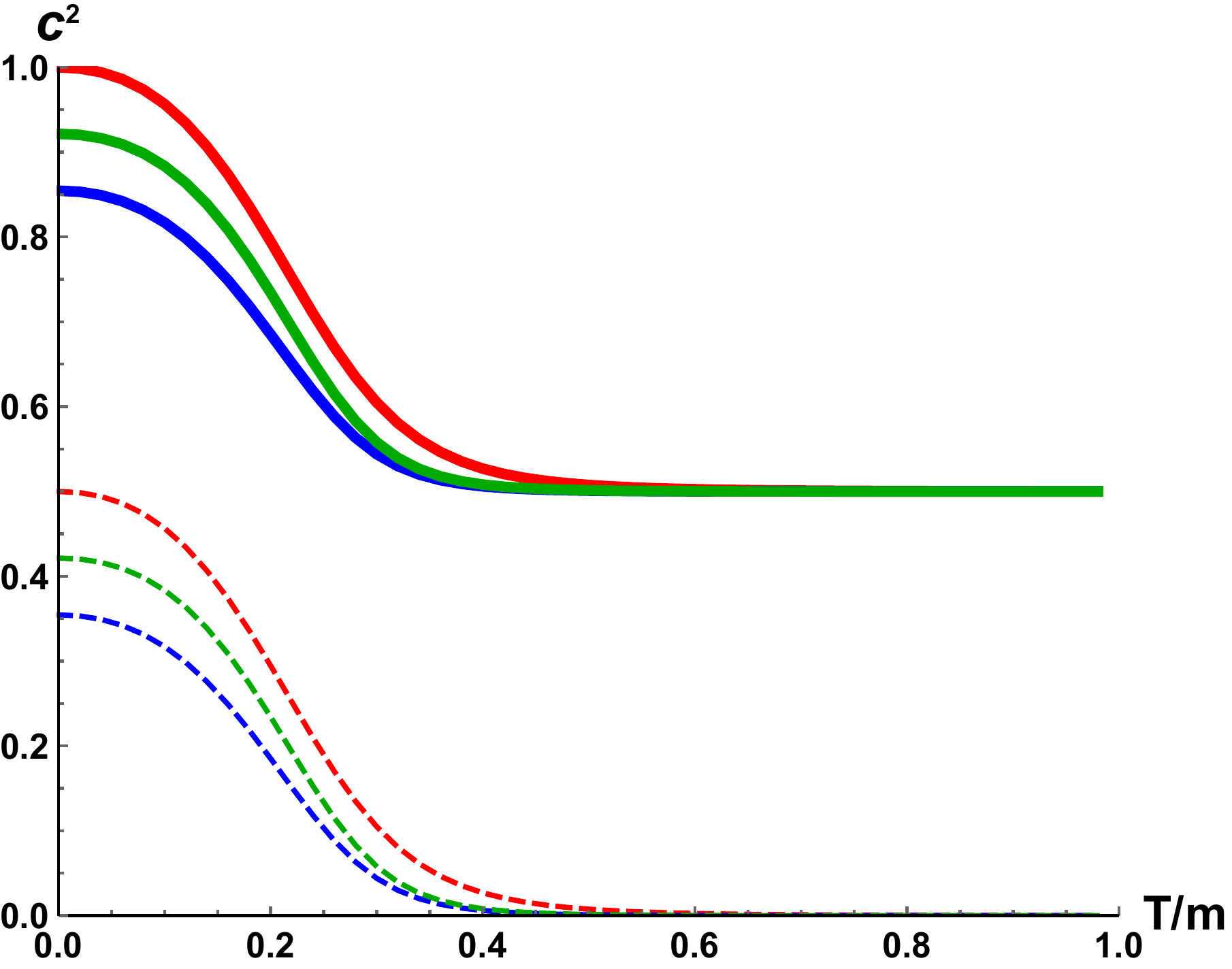}

\vspace{0.3cm}

\includegraphics[width=0.8\linewidth]{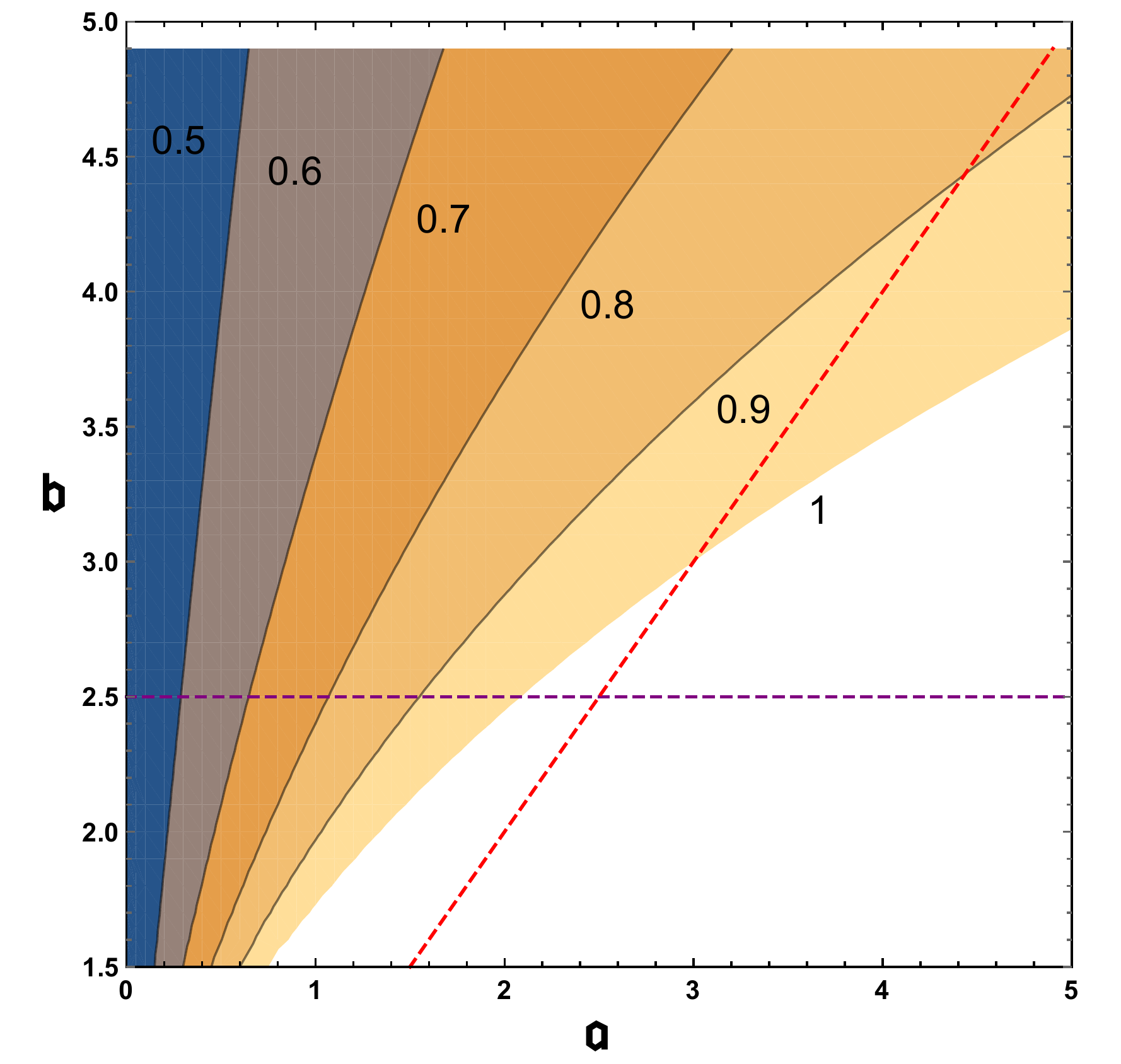}
 \caption{\textbf{Top: }The longitudinal (thick) and transverse (dashed) speeds in function of $T/m$. The specific $(\mathfrak{a},\mathfrak{b})$ are $(3,3),(3,4),(4,4)$ for red, blue and green. \textbf{Bottom: }Values of $c_L^2$ at $T=0$ as extracted from the formula \eqref{speeds2}.  The dashed purple line indicates the region where the phonons are gapless, which happens for $b>5/2$ \cite{Alberte:2017oqx}. Below this line, the formulae in \eqref{speeds2} (and consequently this plot) need not apply. The dashed red line is the potential $V(X,Z)=X^\mathfrak{a}$. More importantly the white region signals the appearance of superluminal longitudinal phonons.}
 \label{figv}
\end{center}
\end{figure}

As already introduced in the previous sections, the linear elastic response directly defines the speed of propagation of transverse and longitudinal sounds in terms of the elastic moduli via eq.\eqref{speeds-0}. Notice that formulae \eqref{speeds-0} are strictly speaking valid only in presence of massless phonons $\omega_{T,L}=c_{T,L}\, k$ and therefore only for $\mathfrak{b}>5/2$. The latter disagreement for  $\mathfrak{b}<5/2$ has been explicitly checked for in \cite{Alberte:2017cch}. The validity of formula \eqref{speeds-0} has been ascertained directly by a direct comparison with the QNMs spectrum for transverse \cite{Alberte:2017oqx,Alberte:2017oqx} and longitudinal \cite{Ammon:2019apj,Baggioli:2019abx} waves. 
Two important results follow:
\begin{itemize}
\item For $\mathfrak{a}=0$ the speed of transverse sound is zero. In addition the speed of longitudinal sound is constant $c_L^2=1/2$ and independent of the power $\mathfrak{b}$. The phase dual to $\mathfrak{a}=0$ is a fluid.
\item At any value of $T/m$ the relation $c_L^2\,=\,\frac{1}{2}\,+\,c_T^2$ holds. This is ensured by conformal symmetry \cite{Esposito:2017qpj} and it is proven by direct computation. 
\end{itemize}
The results for the speeds are shown in Fig.~\ref{figv}. The left panel of Fig.~\ref{figv} shows a typical behaviour of the speeds in function of the dimensionless temperature $T/m$. The speeds exhibit a continuous transition towards the large $T$ values $c_T=0,c_L^2=1/2$. The right panel shows the value of the longitudinal speed at zero $T$, the maximal speed in the system, inside the parameter space. The white region evidentiates the region where the longitudinal speed is superluminal $c_L>1$. Curiously, the speed becomes superluminal in the direction where the dual CFT becomes more and more exotic (auxetic).\\

\begin{figure}[t]
\begin{center}
\includegraphics[width=0.7\linewidth]{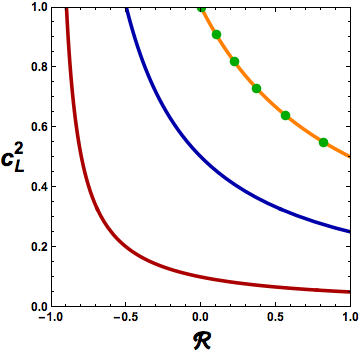}
 \caption{
The longitudinal speed $c_L^2$ in function of the  Poisson ratio $\mathcal{R}$, for various values of $c_e/c$.
The orange line  is for $c_e/c = 1$ and it coincides with the plot obtained for the {\it conformal solid EFT} (shown as green bullets).
The other lines correspond to CFTs the representative values  $c_e/c = 1/\sqrt{2}$ and $1/\sqrt{10}$.  
Notice that the {\it auxetic} behaviour (corresponding to $\mathcal{R}<0$) can be achieved in the CFT case with  $c_e/c<1$,
thus the  {\it conformal solid EFTs } must have $\mathcal{R}>0$ and can't be much auxetic.
Similar plots are obtained for the family of solids with spontaneously broken SI introduced in Section 3, by decreasing the equation of state parameter $w$, which plays a role analogous to $c_e^2/c^2$.
}
 \label{vs}
\end{center}
\end{figure}

Restoring the units in the previous equation is simply done by noticing that $c_{L,\,T}$ are expressed in the units of the universal light-cone speed present in AdS gravity side, $c_e$.
Therefore, restoring the units we recover readily \eqref{SumRule2}. As emphasized in the introduction, this has an  important physical consequence: taking $c_e\ll c $, we obtain a model for a scale invariant material whose sound speeds can be as slow as necessary. For real world applications, notice that $c_e/c$ should be at most of the order $10^{-4}$.

This also makes manifest another important point: in the present construction, the AdS gravity model also shares this small universal speed $c_e$. Therefore, it is clear that our framework has no knowledge of the underlying ultraviolet completion (where eventually the speed of light $c$ plays a key role), including any relation to string compactifications. This is in tune with the view that holography can be used as an effective method as argued before e.g. in \cite{Baggioli:2016oju,Baggioli:2016oqk}.

Let us finish by showing the relation between the Poisson ratio ${\cal R}$ and the longitudinal sound speed in the holographic CFT model,
\be\label{Rce}
{c_L^2} = \frac{c_e^2}{{\cal R}+1}~.
\ee
This makes manifest that in this theory ${\cal R}$ can be negative (auxetic) if $c_e<c$, and the how much auxetic the solid can be depends on how small the ratio $c_e/c$ ie. We show this in Fig.~\ref{vs}. A similar plot can be obtained for the generic EFTs with spontaneously broken scale invariance presented in Section 3, with the equation of state parameter $w$ playing the role of $c_e^2/c^2$.

\section{Conclusions}\label{section:conclusions}

We have studied the possible realization of scale invariance (SI) in the mechanical response of solid materials, considering both the cases for which SI is a spontaneously broken or a manifest symmetry. The latter case takes the form of a nontrivial infrared fixed point (IRFP), and we have studied it using AdS/CFT methods. 

For the spontaneously broken case, it is consistent to assume that the mass-spectrum is gapped and then use Effective Field Theory (EFT) methods to study the gapless phonons as Goldstone bosons of the spontaneously broken spacetime symmetries. 
We have presented a new one-parameter family of EFTs of sponteneously broken SI, which is basically parametrized by the equation of state parameter of the material in the background, $w=p/\rho$. This family includes the previously studied case of the so-called {\it conformal solid} EFT \cite{Esposito:2017qpj} as a special point, however for the generic values it nonlinearly  realizes SI, but not the full conformal group.

We have shown that for both types of realization of SI it is possible to have slow speeds of sound compared to the speed of light, once  $w$ is small. 
More specifically, we found that the longitudinal and transverse sound speeds are related as
\be\label{speedsv}
c_L^2 = w\, c^2 + \frac{2\,(d-2)}{d-1}\,c_T^2
\ee
with $w$ the equation of state parameter of the material and $c$ the speed of light. For small $w$ the two speeds are  small compared to $c$, as needed for potentially realistic applications.

We have constructed manifest SI theories as holographic duals of AdS models. In these case, we also found that \eqref{speedsv} applies.
We have argued that our models can be interpreted as IRFP where in addition to having manifest SI, the theory is also Lorentz invariant with respect to a small `emergent' light-cone speed $c_e$ (so that the theory is actually a full Conformal Field Theory). The symmetries then relate
% $$c_e=\sqrt{w\,(d-1)}\,c$$
 $$w=\frac{1}{d-1}\frac{c_e^2}{c^2}~.$$
Importantly, note that assuming $c_e/c \ll1$ is a consistent possibility, and it immediately leads to slow phonons also in this CFT-like case.
Another interesting consequence of having a full emergent CFT with is that the relation between the speeds and the elastic moduli change by $O(1)$ factor. For instance the transverse sound speed becomes $c_T^2 = \frac{d-1}{d} \,\frac{\mathcal{G}}{\rho_m} $ instead of the usual form $\frac{\mathcal{G}}{\rho_m} $, with $G$ the shear modulus and $\rho_m$ the mass density. Thus, a deviation from the  usual formula $\frac{\mathcal{G}}{\rho_m} $ could be used as a smoking gun of the emergent SI and Lorentz in the material. 
The measurement of elastic properties in some cuprates has been recently done in Ref. \cite{Sahu_2019}. 
Our work motivates further investigation in this direction.

In our models with manifest SI, we have used standard holographic methods to compute several  elastic response parameters: the elastic moduli, the Poisson ratio, the propagation speed of the phonons in function of the various parameters of the model.  
We find that the maximally auxetic solids (most negative Poisson ratio) can arise only for $w\ll 1$, that is, for slow sound speeds.
Also, the temperature dependence of the different features suggest once more that these holographic models seem to interpolate between a fluid phase to a solid phase by decreasing temperature. The crossover is continuous and very analogous to what happens to certain extent in glasses and amorphous materials. The behaviour of the vibrational modes in these holographic systems has already produced important developments in the study of the latter \cite{Baggioli:2018qwu,baggioli2018soft}.

An obvious extension of this work, which is currently under investigation, amounts to generalize these results to the non-linear regime up to arbitrarily large deformations. An EFT description has been recently introduced in \cite{Alberte:2018doe}. We plan to report in a forthcoming work \cite{next} the analysis of nonlinear elasticity in  scale invariant system,  comparing also the spontaneously broken {\it vs.} manifest cases.

Finally we hope that this and related works stimulate further experimental investigation towards the mechanical properties and the phonons dynamics in quantum critical situations and scale invariant systems. Preliminary interesting studies have been presented in \cite{setty2019glass,ishii2019glass}; more has definitely to come.

\section*{Acknowledgements}
We thank L.Alberte, M.Ammon, T.Andrade, A. Esposito, C, Hoyos,  A. Jimenez Alba, A. Nicolis, R. Penco, A.Zaccone, K.Trachenko, Jan Zaanen, Weijia Li and Milan Allan for useful discussions.
OP thanks the organizers of the workshop ``Effective theories of quantum phases of matter'' in Nordita, which stimulated the completion of this work.
We acknowledge the support of the Spanish Agencia Estatal de Investigacion through  
the grant FPA2017-88915-P and the Severo Ochoa Excellence grants SEV-2016-0597 and SEV-2016-0588, as well as from the DURSI through grant 2017-SGR-1069.MB thanks Jiao Tong Shangai University, TDLee Institute, ITP Beijing and DUT Dalian for the kind hospitality during the completion of this work.

\appendix 

\section{More on solid EFT in $d$ dimensions}\label{app}
In order to compute the speeds of the phonon modes we need to calculate the quadratic action on the perturbations around the background of the fields $\Phi^I = \alpha\, (x^I + \pi^I)$ which give us 
\begin{equation}
\mathcal{I}_{IJ} \, = \, \alpha^2 (\delta_{IJ}+ \partial_I \pi_J + \partial_J \pi_I + \partial^{\,\mu} \pi_I \,\partial_\mu \pi_J). 
\end{equation}
The most relevant expressions up to quadratic order are summarized below
\begin{align}
\mbox{Tr} (\mathcal{I}^{\,n} )\, = \,& \alpha^{2n} \big(  (d-1) - n \,\dot{\pi_i}^2 + 2\,n\, \partial_i\, \pi^L_i \nonumber\\&+ n\,(2n-1)\, ( \partial_i\, \pi^L_i)^2 + n^2\,  ( \partial_i \,\pi^T_j)^2 \big),
\end{align}
\begin{equation}
Z \, = \, \alpha^{2(d-1)} \left( 1 + 2\, \partial_i\,\pi^L_i  - \dot{\pi_i}^2 +  ( \partial_i \,\pi^L_i)^2  \right),
\end{equation}
\begin{equation}
x_n \equiv \frac{\mbox{Tr} (\mathcal{I}^{\,n}) }{Z^{\frac{n}{d-1}}} = d-1 + n^2\, ( \partial_i \,\pi^T_j)^2 + \frac{2(d-2)}{d-1}\,n^2 \,( \partial_i \,\pi^L_i)^2,
\end{equation}
where we have split  the perturbation into longitudinal and transverse modes
\begin{equation}
\partial_i\, \pi^T_i \,=\, 0 \quad , \quad \, \partial^{\,}_{[i} \, \pi^L_{j]} \,=\, 0.
\end{equation}
The action at second order is then 
\begin{equation}
\delta S^{(2)} = - \int d^d x \left(-\,N\, \dot{\pi_i}^2 + c_L^2 \,( \partial_i \,\pi^L_i)^2 +  c_T^2 \,( \partial_i\, \pi^T_j)^2 \right) 
\end{equation}
where 
\begin{equation}
N = Z\,V_Z\,,
\end{equation}
\begin{equation}
c_L^2 \, = \, 1 + \frac{2\, V_{ZZ}\, Z}{V_Z} + \frac{2(d-2)}{d-1} \, c_T^2\,,
\end{equation}
\begin{equation}
c_T^2 \, =\, \sum_{n=1}^{d-2} \, \frac{n^2 \, V_{x_n}}{Z \, V_Z}.
\end{equation}
We would like to relate this to the bulk and shear moduli. The stress-energy tensor is 
\begin{equation}
T_{\mu\nu} \, = \, - \frac{2}{\sqrt{-g}}\,\frac{\delta S}{\delta g^{\mu \nu}} \, = \, - \eta_{\mu\nu}\, V + 2\, \frac{\partial V}{\partial \mathcal{I}^{IJ}} \, \partial_\mu \Phi^I \, \partial_\nu \Phi^J.
\end{equation}
Our potential is a function of $Z$ and $x_n$, so 
\begin{align}
&\frac{\partial V}{\partial \mathcal{I}^{IJ}} \, = \, \frac{\partial Z}{\partial \mathcal{I}^{IJ}}\, \frac{\partial V}{\partial Z } \,\nonumber \\&+  \sum_{n=1}^{d-2} \Big( \frac{\partial \mbox{Tr} (\mathcal{I}^{\,n}) }{\partial \mathcal{I}^{IJ}}\,\frac{1}{Z^{n/d-1}}\, -\, \frac{\partial Z}{\partial \mathcal{I}^{IJ}} \, \frac{n}{d-1} \,\frac{x_n}{Z} \Big) \frac{\partial V}{\partial x_n}~.
\end{align}
Let's start computing the shear modulus. The shear strain changes our background to 
\begin{equation}
\Phi^I = x^I + \varepsilon^I_k x^k
\end{equation}
where we can take $\varepsilon^I_J \, = \, \varepsilon^J_I$ with no loss of generality. We assume that $\varepsilon^j_i\neq0$ for $i\neq j$ and look at the component $T_{ij}$ at first order in $\varepsilon^j_i$, and exptract the shear modulus comparing with Eq.\eqref{Tlin}. Notice that the term $\frac{\partial Z}{\partial \mathcal{I}^{IJ}}$ cancels with $\partial_i \Phi^I \partial_j \Phi^J$. To check this, first we make the derivative of $Z$ with respect to $\mathcal{I}^{IJ}$ using Jacobi's formula
\begin{equation}
\frac{\partial Z}{\partial \mathcal{I}^{IJ}} = \mbox{adj}^T(\mathcal{I})_{IJ}.
\end{equation}
Contracting this with
\begin{equation}
\partial_i \Phi^I \partial_j \Phi^J = (\delta^I_i \delta^J_j + \delta^I_i\varepsilon^J_j + \delta^J_j \varepsilon^I_i)\,\alpha^2 \, + \, \mathcal{O}(\varepsilon^2).
\end{equation}
At linear order we find that $\mbox{adj}^T(\mathcal{I})_{ij} = -2\,\varepsilon_{ij}\, \alpha^{2d-4}$ (with $i\neq j$) and $\mbox{adj}^T(\mathcal{I})_{ii} = \alpha^{2d-4} (1+\mathcal{O}(\varepsilon))$. Therefore
\begin{align}
&\frac{\partial Z}{\partial \mathcal{I}^{IJ}} \, (\delta^I_i \delta^J_j + \delta^I_i\varepsilon^J_j + \delta^J_j \varepsilon^I_i)\,\alpha^2 \nonumber \\ &= \mbox{adj}^T(\mathcal{I})_{ij} \, \alpha^2 \,+\, 2 \, \alpha^{2(d-1)}\,\varepsilon_{ij}\,=\,0.
\end{align}
The only non-zero term is then
\begin{equation}
T_{ij} \, = \, 2 \,\sum_{n=1}^{d-2} \frac{\partial V}{\partial x_n} \,\frac{\partial \mbox{Tr} ((\mathcal{I}^{KL})^n) }{\partial \mathcal{I}^{IJ}}\, \frac{1}{Z^{n/d-1}}\, \partial_i \Phi^I\, \partial_j \Phi^J.
\end{equation}
For the derivative of the traces,\\ $\mbox{Tr} (\mathcal{I}^n) = \mathcal{I}^{I_1 \, I_2} \,\mathcal{I}^{I_2 \,I_3}\,\dots \,\mathcal{I}^{I_{n-1} \, I_n}\,\mathcal{I}^{I_n \, I_1}$,  one finds 
$$
\frac{\partial \mbox{Tr}{(\mathcal{I})} }{\partial \mathcal{I}^{IJ}} = \delta_{IJ} \,,\qquad
\frac{\partial \mbox{Tr}{(\mathcal{I}^2)} }{\partial \mathcal{I}^{IJ}} = 2 \,\mathcal{I}_{IJ}\,,
$$
and 
\begin{equation}
\frac{\partial \mbox{Tr} (\mathcal{I}^n) }{\partial \mathcal{I}^{IJ}} \, = \, n \,\mathcal{I}^{I \, I_3} \,\mathcal{I}^{I_3 \,I_4}\,\dots \,\mathcal{I}^{I_{n-1} \, I_n}\,\mathcal{I}^{I_n \, J}
\end{equation}
for $n>2$, where we have used the cyclic property of the trace.
Since $\mathcal{I}^{IJ} = \alpha^2 (\delta_{IJ}+ \varepsilon^I_J + \varepsilon^J_I)$, finally we can find that at linear order 
\begin{equation}
T_{ij} \, = \, 4\,\varepsilon_{ij} \,\sum_{n=1}^{d-2} \,n^2 \,\frac{\partial V}{\partial x_n} \,= \,2 \, \varepsilon_{ij}\, \mathcal{G}.
\end{equation}
For the bulk modulus we consider a purely volume deformation (zero shear), which can be parametrized as
\be\label{kappa}
\alpha = 1+\frac{\kappa}{d-1}
\ee 
where $\kappa$ is the bulk strain, and we look at $T_{ii}$. \\
Notice that $V_{x_n}$ doesn't appear here as we can easily check using $\partial_i \Phi^I\, \partial_i \Phi^J = \alpha^2 \delta^I_{i}\,\delta^J_i$
\begin{equation}
\left( \frac{\partial \mbox{Tr} (\mathcal{I}^{\,n}) }{\partial \mathcal{I}^{ii}}\,\frac{1}{Z^{n/d-1}}\, -\, \frac{\partial Z}{\partial \mathcal{I}^{ii}} \, \frac{n}{d-1} \,\frac{x_n}{Z} \right) \alpha^2 \,=\, 0.
\end{equation}
Therefore we arrive to
\begin{equation}\label{Tii}
T_{ii} \,=\, - V + 2\, Z\, V_Z 
\end{equation}
and from the equation above we can already find the important result 
\be\label{r+p}
\rho + p  = 2 \,Z\, V_Z~.
\ee

Finally, using the definition of the bulk modulus \eqref{Kdef} together with $\mathcal{V}\propto \alpha^{1-d}$  and \eqref{kappa}, one arrives at
\begin{equation}
\mathcal{K} \equiv - \mathcal{V}\,\frac{dp}{d\mathcal{V}} = \,\frac{dp}{d\kappa}\,=\,\frac{dT_{ii}}{d\kappa}
\end{equation}
From \eqref{Tii}, then, one finds
\begin{equation}
\mathcal{K} \,=\, 2 \,Z\, V_Z\, +\,4\, Z^2\, V_{ZZ} = 4 \,Z^{3/2} \,\partial_Z \left( \sqrt{Z}\, V_Z \right).
\end{equation}

It is also possible to rewrite, the bulk modulus in terms of  the equation of state of the solid, understood as the functional dependence of the pressure on the energy density, that is $P(\rho)|_{{}_{\Box}}$, by changing only the density -- that is at zero shear strain. Note that
\begin{equation}
\mathcal{K}\,=\,\frac{dp}{d\rho}\,\frac{d\rho}{dZ}\,\frac{dZ}{d\kappa} \, = \, \frac{dp}{d\rho}\,2\,Z\,V_Z,
\end{equation}
where we use $\rho = V$. Therefore we find
\begin{equation}
\frac{\mathcal{K}}{\rho+p}\,=\, \frac{dp}{d\rho} \Big|_{{}_{\Box}}~,
\end{equation}
which leads to \eqref{cs-general}. The subscript $|_{{}_{\Box}}$ stands to recall that the derivative is at vanishing shear deformation.

\bibliographystyle{apsrev4-1}

\bibliography{NLE}

\end{document}